\documentclass[11pt]{article}

\usepackage[utf8]{inputenc}

\usepackage{
amsmath,amssymb,amsfonts,amsthm}
\usepackage{bm}
\usepackage{colonequals}
\usepackage{comment}
\usepackage{epsfig} \usepackage{latexsym,nicefrac,bbm}
\usepackage
{xspace}
\usepackage{color,fancybox,graphicx,subfigure,fullpage}
\usepackage[top=1.25in, bottom=1.25in, left=1in, right=1in]{geometry}
\usepackage{tabularx}
\usepackage{hyperref}
\usepackage{cleveref}
\usepackage[boxruled,linesnumbered]{algorithm2e}
\usepackage{siunitx}
\usepackage{mathtools}

\usepackage{mathrsfs}

\usepackage{multicol}
\usepackage{enumitem}
\usepackage{float}
\usepackage{tikz}
\usepackage{framed}

\makeatletter
\newcommand{\vast}{\bBigg@{4}}
\newcommand{\Vast}{\bBigg@{5}}
\makeatother

\newtheorem{theorem}{Theorem}[section]

\newtheorem{definition}[theorem]{Definition}
\newtheorem{lemma}[theorem]{Lemma}
\newtheorem{remark}[theorem]{Remark}
\newtheorem{proposition}[theorem]{Proposition}
\newtheorem{corollary}[theorem]{Corollary}

\DeclareMathOperator{\arcsinh}{arcsinh}

\newcommand{\WRank}{\mathrm{WRank}}
\newcommand{\Inv}{\mathrm{Inv}}
\newcommand{\TFrame}{\mathrm{TFrame}}
\newcommand{\Orthog}{\mathrm{Orthog}}
\newcommand{\WPHP}{\mathrm{WPHP}}
\newcommand{\SRank}{\mathrm{SRank}}
\newcommand{\Rigid}{\mathrm{Rigid}}
\newcommand{\ncABP}{\mathrm{ncABP}}
\newcommand{\CPRank}{\mathrm{CPRank}}
\newcommand{\NormWRank}{\mathrm{NormWRank}}

\newcommand{\imag}{\mathrm{i}}

\newcommand{\CP}{\mathrm{CP}}

\newcommand{\bC}{\mathbb{C}}

\newcommand{\bE}{\mathbb{E}}
\newcommand{\bF}{\mathbb{F}}

\newcommand{\bN}{\mathbb{N}}

\newcommand{\bR}{\mathbb{R}}

\DeclareSymbolFont{bbold}{U}{bbold}{m}{n}
\DeclareSymbolFontAlphabet{\mathbbold}{bbold}
\newcommand{\One}{\mathbbold{1}}

\newcommand{\sign}{\mathrm{sign}}

\newcommand{\rank}{\mathrm{rank}}
\newcommand{\supp}{\mathrm{supp}}
\newcommand{\signrank}{\mathrm{srank}}

\DeclareMathOperator{\tr}{tr}

\renewcommand{\epsilon}{\varepsilon}
\renewcommand{\Tilde}{\widetilde}
\renewcommand{\hat}{\widehat}

\newcommand\numberthis{\addtocounter{equation}{1}\tag{\theequation}}

\newcommand\zzero{\boldsymbol{0}}

\def\calI{\mathcal{I}}

\def\calL{\mathcal{L}}

\def\calR{\mathcal{R}}
\def\calS{\mathcal{S}}

\def\calZ{\mathcal{Z}}

\renewcommand\AA{\bm{A}}
\newcommand\BB{\bm{B}}

\newcommand\EE{\bm{E}}
\newcommand\FF{\bm{F}}
\newcommand\GG{\bm{G}}

\newcommand\II{\bm{I}}

\newcommand\MM{\bm{M}}
\newcommand\NN{\bm{N}}
\newcommand\OO{\bm{O}}

\renewcommand\SS{\bm{S}}
\newcommand\TT{\bm{T}}

\newcommand\XX{\bm{X}}
\newcommand\YY{\bm{Y}}
\newcommand\ZZ{\bm{Z}}

\newcommand\sB{\mathsf{B}}
\newcommand\sC{\mathsf{C}}

\newcommand\sE{\mathsf{E}}

\newcommand\sM{\mathsf{M}}

\newcommand\sP{\mathsf{P}}

\newcommand\sU{\mathsf{U}}
\newcommand\sV{\mathsf{V}}

\newcommand\sX{\mathsf{X}}
\newcommand\sY{\mathsf{Y}}
\newcommand\sZ{\mathsf{Z}}
\newcommand\st{\mathsf{t}}

\newcommand\sv{\mathsf{v}}
\newcommand\se{\mathsf{e}}
\newcommand\ssv{\bm{\mathsf{v}}}
\newcommand\sse{\bm{\mathsf{e}}}

\def\aa{\bm{a}}
\newcommand\bb{\bm{b}}
\newcommand\cc{\bm{c}}

\newcommand\ee{\bm{e}}

\newcommand\pp{\bm{p}}

\newcommand\uu{\bm{u}}
\newcommand\vv{\bm{v}}

\newcommand\xx{\bm{x}}

\newcommand\yy{\bm{y}}
\newcommand\zz{\bm{z}}

\newcommand\aalpha{\boldsymbol{\alpha}}

\usepackage{authblk}

\title{The sum-of-squares complexity of linear-algebraic principles\thanks{This research was partially supported by ONR Award N00014-24-12611.}}
\author[1]{Dmitriy Kunisky\thanks{Email: \texttt{kunisky@jhu.edu}.}}
\author[2]{Xifan Yu\thanks{Email: \texttt{xifan.yu@yale.edu}.}}
\affil[1]{Department of Applied Mathematics \& Statistics, Johns Hopkins University}
\affil[2]{Department of Computer Science, Yale University}

\begin{document}

\pagenumbering{roman}

\maketitle

\thispagestyle{empty}

\begin{abstract}
We study the degree needed for sum-of-squares (SoS) proofs of several basic principles of real linear algebra.
For any $n\times n$ matrix $\AA$ of rank greater than $r$, the Weak Rank Principle states that the equations $\sX\sY^\top=\AA$ with $n\times r$ matrix variables $\sX,\sY$ have no solution.
We show that this principle requires degree exactly $2r+2$ to prove.
On the other hand, for $\AA=\II_n$, setting $\sX = \sY$ gives the Tight Frame Principle, which has a degree-$4$ proof.
For $n\times n$ matrices, proving that a right inverse is also a left inverse requires degree exactly $2n+2$, while proving that orthonormal rows imply orthonormal columns again has a degree-4 proof.
Using our lower bounds, we also derive related SoS degree lower bounds for polynomial systems encoding sign rank, matrix rigidity, small-width homogeneous non-commutative algebraic branching programs, and tensor CP rank.
Lastly, we show that quantitative bounds on the Frobenius norms of the factors in the Weak Rank Principle can lower the degree required to refute $\sX\sY^\top=\II_n$, but sufficiently large bounds still preserve the exact degree $2r+2$ from the unbounded case, giving a quantitative version of the Archimedean property arising in classical results on SoS proofs.
\end{abstract}

\clearpage

\pagestyle{empty}

\tableofcontents

\clearpage

\pagestyle{plain}
\pagenumbering{arabic}

\allowdisplaybreaks{

\section{Introduction}

Elementary linear algebra gives short proofs of many statements about rank and matrix inverses that can be viewed as the infeasibility of certain systems of polynomial equations.
Our goal in this paper is to study how much of this reasoning the sum-of-squares (SoS) proof system~\cite{BS-2014-SOSQuest,Laurent-2009-SOS} can carry out at low degree, when the polynomial equations are posed over $\mathbb{R}$.
Our results show that closely related linear-algebraic principles can in fact require very different SoS degrees, and that these differences lead to limitations on proving lower bounds for several algebraic complexity measures.

Our main result determines the exact SoS degree required to show the \emph{Weak Rank Principle}: if $\AA$ has rank greater than $r$, then $\AA$ cannot be written as $\sX\sY^\top$ with $\sX,\sY$ having $r$ columns (in our notation, bold letters denote constant matrices while sans serif letters denote variables in the polynomial systems we study).
Every such instance requires SoS degree exactly $2r+2$ (Theorem~\ref{thm:WRank-thm}).
For the identity matrix, identifying the two factors $\sX = \sY$ gives the \emph{Tight Frame Principle}, which says that $n>r$ orthonormal vectors cannot lie in $\bR^r$.
This principle, in contrast fo the Weak Rank Principle, has a degree-$4$ SoS proof (Theorem~\ref{thm:TFrame-thm}).
The same distinction appears for $n\times n$ matrices: proving that a right inverse is also a left inverse requires degree exactly $2n+2$, while proving that a matrix with orthonormal rows also has orthonormal columns has a degree-4 proof (Theorems~\ref{thm:Inv-thm} and~\ref{thm:Orth-thm}).
We also give applications to other polynomial systems of interest in the literature, using our Weak Rank Principle lower bound to obtain related SoS lower bounds for polynomial systems expressing small sign rank, low matrix rigidity, the existence of a small-width homogeneous non-commutative algebraic branching program, and small tensor CP rank.

Finally, we examine the effect on this SoS complexity of bounding the norms of the factors.
The Tight Frame Principle equations directly fix the squared Frobenius norm of $\sX$ at $n$.
We find that, for $\AA=\II_n$, adding an explicit bound on $\|\sX\|_F^2+\|\sY\|_F^2$ to the Weak Rank Principle equations can analogously reduce the degree needed for a refutation.
For sufficiently large norm budgets, however, the exact lower bound $2r+2$ persists (Theorems~\ref{thm:norm-constraints-upper-bound} and~\ref{thm:norm-constraints-lower-bound}).
Thus, this implicit boundedness is a plausible mechanism for why the Tight Frame Principle is so much easier for SoS to derive than the Weak Rank Principle.
Further, the specific size of the norm bound imposed in the modified Weak Rank Principle matters, yielding an interesting quantitative version of the \emph{Archimedean} boundedness property appearing in literature on more abstract existence results for SoS-type proofs and certificates.

\subsection{Sum-of-Squares Proof System}

Algebraic proof systems certify that polynomial equations have no common solution by exhibiting a polynomial identity~\cite{PT-2016-AlgebraicProofComplexity}.
For instance, Hilbert's Nullstellensatz says that a system $p_1=\cdots=p_s=0$ has no solution over an algebraically closed field (such as $\mathbb{C}$) if and only if $1=\sum_i h_i p_i$ for suitable polynomials $h_i$.
Over $\mathbb{R}$, the sum-of-squares proof system further allows a certificate to use nonnegativity as well as the given equations.
For example, the equation $x^2+1=0$ is feasible over $\mathbb{C}$ but is refuted over $\mathbb{R}$ by the identity $-1=x^2-(x^2+1) \geq 0$.
Stengle's Positivstellensatz guarantees a certificate of this form for every infeasible system of real polynomial equations~\cite{stengle1974nullstellensatz}.

The SoS proof system, introduced in this form by Grigoriev and Vorobjov~\cite{grigoriev2002complexity}, measures the degree needed in such a certificate.
At a fixed degree, searching for a certificate amounts to a semidefinite feasibility problem, where sums of squares are represented by positive semidefinite Gram matrices, and the polynomial identity imposes linear constraints on their entries and on the coefficients of the multipliers~\cite{Laurent-2009-SOS}.
The degree of SoS proof therefore controls the size of this formulation.
We use the following convention, in which a collection of polynomials denotes the system obtained by setting each polynomial equal to zero.

\begin{definition}[SoS Refutation]
    Let $S = \{p_1, \dots, p_s\} \subseteq \bR[\sX]$ be a collection of polynomials. A degree-$D$ SoS refutation of the system $S$ of polynomials is a polynomial identity
    \begin{align*}
        -1 &= g + \sum_{i=1}^s h_ip_i,
    \end{align*}
    where $h_i \in \bR[\sX]$ and $\deg(h_ip_i) \le D$ for every $1 \le i \le s$, and $g \in \bR[\sX]$ is a sum-of-squares polynomial of degree at most $D$.
    In particular, an SoS refutation of $S$ gives a proof that the system $S$ has no common solutions.
\end{definition}
\noindent
We also consider the following notion of SoS proof of polynomial identity.

\begin{definition}[SoS Proof]
    Let $S = \{p_1, \dots, p_s\}$ and $T = \{q_1, \dots, q_t\} \subseteq \bR[\sX]$ be two collections of polynomials. A degree-$D$ SoS proof of the system $T$ of polynomials given the system $S$ of polynomials is a polynomial identity
    \begin{align*}
        \sum_{i=1}^t (1 + k_i)q_i^2 + g &= \sum_{i=1}^s h_ip_i,
    \end{align*}
    where $h_i \in \bR[\sX]$ and $\deg(h_ip_i) \le D$ for every $1 \le i \le s$, $k_i$ is a sum-of-squares polynomial and $\deg((1+k_i)q_i^2) \le D$ for every $1 \le i \le t$, and $g \in \bR[\sX]$ is a sum-of-squares polynomial of degree at most $D$.
    In particular, an SoS proof of $T$ given $S$ gives a proof that any solution to the system $S$ is a solution to the system $T$.
\end{definition}

We call a \emph{principle} the formal proposition that a certain polynomial inequality or implication of polynomial systems either holds or does not hold.
For a principle $P$, we write $\vdash_D P$ to mean that $P$ has an SoS proof of degree at most $D$.
If $P$ asserts that a polynomial system $S$ is infeasible, this means a degree-$D$ SoS refutation of $S$.
If $P$ asserts that every solution of a polynomial system $S$ also solves a system $T$, this means a degree-$D$ SoS proof of $T$ given $S$.

Throughout, degree means total polynomial degree, including the multipliers in the displayed identities.
All indeterminates range over $\bR$, and we impose Boolean constraints such as $x^2-x=0$ only when stated explicitly.
Our lower bounds refer to the given polynomial encodings and to the above notions of proof and refutation.

We use polynomial substitutions to transfer SoS proofs between systems.
Substituting polynomials of degree at most $k$ for the variables preserves sums of squares and increases the proof degree by at most a factor of $k$.
We also will consider restricting some variables to real constants (equivalently, substituting constants for those variables).
We note that affine substitutions and restrictions obtained by fixing variables to real constants do not increase degree.

\subsection{Weak Rank Principle, Inversion Principle, and Related Principles}

Let $n > r$ be two positive integers. We consider the following linear algebraic principles over the reals.

\begin{definition}[Weak Rank Principle]
    Let $\sX, \sY$ be two $n \times r$ matrices of real indeterminates. The Weak Rank Principle, denoted $\WRank_{r}^n$, states that for an arbitrary matrix $\AA \in \bR^{n \times n}$ of rank at least $r + 1$, the system
    \begin{align*}
        \sX\sY^\top = \AA
    \end{align*}
    is infeasible. We denote the infeasible system parameterized by $\AA \in \bR^{n \times n}$ by $\WRank_{r}^n(\AA)$.
\end{definition}

\begin{definition}[Inversion Principle]
    Let $\sX, \sY$ be two $n \times n$ matrices of real indeterminates. The Inversion Principle, denoted $\Inv_n$, states that
    \begin{align*}
        \sX\sY^\top = \II_n \Longrightarrow \sY^\top \sX = \II_n.
    \end{align*}
\end{definition}

We also study two closely-related principles obtained from replacing $Y$ with the same matrix $X$ in WRank and Inv.

\begin{definition}[Tight Frame Principle]
    Let $\sX$ be an $n \times r$ matrix of real indeterminates. The Tight Frame Principle, denoted $\TFrame_r^n$, states that the system
    \begin{align*}
        \sX\sX^\top = \II_n
    \end{align*}
    is infeasible.
\end{definition}

\begin{remark}
    The Tight Frame Principle may be viewed as the real analogue of the Oddtown Principle \cite[Section~3.1.1]{BBP95}, which states that $\sX\sX^\top = I_n$ has no solutions for an $n \times r$ matrix $\sX$ with entries in $\bF_2$.
\end{remark}

\begin{definition}[Orthogonality Principle]
    Let $\sX$ be an $n \times n$ matrix of real indeterminates. The Orthogonality Principle, denoted $\Orthog_n$, states that the system
    \begin{align*}
        \sX\sX^\top = \II_n \Longrightarrow \sX^\top \sX = \II_n.
    \end{align*}
\end{definition}
\noindent
We also note that WRank is related to the classical Weak Pigeonhole Principle.

\begin{definition}[Weak Pigeonhole Principle]
    Let $\sX$ be an $n \times r$ matrix of real indeterminates. The Weak Pigeonhole Principle, denoted $\WPHP_r^n$, states that the system
    \begin{align*}
        \sX_{i,j}^2 &= \sX_{i,j}, \quad && \forall\, (i,j) \in [n] \times [r]\\
        \sX_{i,j}\sX_{k,j} &= 0, \quad && \forall\, j \in [r], 1 \le i < k \le n\\
        \sum_{j=1}^r \sX_{i,j} &= 1, \quad && \forall\, 1 \le i \le n
    \end{align*}
    is infeasible.
\end{definition}

\begin{remark}[Boolean Weak Rank Principle]\label{rem:boolean-weak-rank}
    Adding the Boolean equations $\sX_{i,k}^2=\sX_{i,k}$ and $\sY_{i,k}^2=\sY_{i,k}$ to $\sX\sY^\top=\II_n$ gives a system with an SoS refutation of degree at most $8$, independent of $n$ and $r$.
    To see this, set $\sC=\sY^\top\sX$.
    Modulo the Boolean equations, each monomial $\sY_{i,k}\sX_{i,\ell}\sY_{j,\ell}\sX_{j,k}$ equals its square, so
    \begin{align*}
        &\sum_{k=1}^r (\sC_{k,k}-1)^2
        +\sum_{\substack{k,\ell\in[r]\\k\ne\ell}}\sum_{i,j=1}^n
        (\sY_{i,k}\sX_{i,\ell}\sY_{j,\ell}\sX_{j,k})^2 \equiv \tr(\sC^2)-2\tr(\sC)+r.
    \end{align*}
    The Boolean reductions use degree at most $8$.
    The identities $\tr(\sC)=\tr(\sX\sY^\top)$ and $\tr(\sC^2)=\tr((\sX\sY^\top)^2)$ then make the right-hand side congruent to $r-n$ modulo $\sX\sY^\top-\II_n$, using degree at most $4$.
    Dividing by $n-r$ gives the claimed refutation.
    Here the matrix equations use real arithmetic.
    Encodings of matrix multiplication over $\bF_2$ instead compute sums modulo $2$, so the finite-field lower bounds discussed below concern different systems.
\end{remark}

We also have the following reductions between the stated principles in the SoS proof system, which we prove in Appendix~\ref{app:proofs-of-reductions}.

\begin{proposition}\label{prop:proof-system-reduction}
    Let $n > r$ be positive integers and let $\AA \in \bR^{n \times n}$ have rank at least $r+1$.
    Then, the following implications hold:
    \[
        {\vdash_D\WRank_r^n(\AA)}
        \Longrightarrow {\vdash_D\WRank_r^n(\II_n)}
        \Longrightarrow {\vdash_D\TFrame_r^n}
        \Longrightarrow {\vdash_D\WPHP_r^n}.
    \]
\end{proposition}

\begin{proposition}\label{prop:inversion-orthogonal-reduction}
    Let $n$ be a positive integer.
    Then, the following implication holds:
    \[
        {\vdash_D\Inv_n} \Longrightarrow {\vdash_D\Orthog_n}.
    \]
\end{proposition}

\begin{proposition}\label{prop:inversion-rank-reduction}
    Let $r \ge 2$ be an integer.
    Then, the following implication holds for $D \ge 4$:
    \[
        {\vdash_D\Inv_{r}} \Longrightarrow {\vdash_D\WRank_r^{r+1}(\II_{r+1})}.
    \]
\end{proposition}

\subsection{Related Work}

Galesi, Grochow, Pitassi, and She~\cite[Theorem~3.4]{galesi2023algebraic} proved that refuting $\WRank_r^n(\II_n)$ in Polynomial Calculus requires degree at least $r/2+1$ over any field, with or without Boolean axioms.
Their proof reduces the functional pigeonhole principle to the rank equations and uses Razborov's degree lower bound~\cite{razborov1998lower}.
They also obtain a degree lower bound of $n/2+1$ for the Inversion Principle with Boolean axioms~\cite[Corollary~3.5]{galesi2023algebraic}.
This reduction alone does not give an SoS lower bound: the pigeonhole system defined above has a degree-$4$ SoS refutation, as follows from Proposition~\ref{prop:proof-system-reduction} and Theorem~\ref{thm:TFrame-thm}.
Our results instead establish exact SoS degrees for the rank and inversion principles with unrestricted real variables.

Garl{\'\i}k, Gryaznov, Ren, and Tzameret~\cite{garlik2026weak} proved exponential size lower bounds for Polynomial Calculus Resolution (PCR) over $\bF_2$ for the algebraic, perfect matching, and bamboo-tree encodings of the Weak Rank Principle.
They used these bounds to construct proof complexity generators and to prove hardness results for statements asserting circuit lower bounds.
Our applications follow the same general strategy of reducing rank equations to equations for more complicated representations.
In particular, we use their recursive encoding of homogeneous non-commutative algebraic branching programs.
Our bounds measure SoS degree over the reals, where the distinction between unrestricted variables and Boolean variables is essential (see Remark~\ref{rem:boolean-weak-rank}).

Garl{\'\i}k, Gryaznov, Lu, Santhanam, and Tzameret~\cite{garlik2026meta} introduced a Tensor Rank Principle based on CP decomposition and proved exponential PCR size lower bounds over $\bF_2$ for diagonal tensor instances.
They also used this principle to study the proof complexity of algebraic circuit lower bounds.
Our tensor result gives SoS degree lower bounds for the real CP decomposition equations, with an explicit dependence on the target rank, tensor order, and dimensions.

\subsection{AI Acknowledgement}

The authors were able to establish a degree-$4$ SoS lower bound for the Weak Rank Principle using an explicit degree-$4$ pseudoexpectation construction. The authors prompted GPT-5.6 Sol Pro to generalize the pseudoexpectation construction to higher degree, but instead obtained an SoS lower bound proof, matching the determinantal upper bound, that does not involve the construction of an explicit pseudoexpectation operator. The original proof provided by GPT 5.6-Sol Pro was stated in the language of algebraic geometry, which was later simplified by the authors and is presented in the current elementary proofs of Theorem~\ref{thm:WRank-thm}, Proposition~\ref{prop:vanishing-top-degree}, and Proposition~\ref{prop:vanishing-structure}.

The authors came up with the question of studying the quantitative effect of imposing an Archimedean property by adding norm constraints to the Weak Rank Principle, and observed that if the norm constraint is too strict, then the degree-$4$ SoS proof for the Tight Frame Principle can be adapted to a degree-$4$ SoS proof for the Norm-Constrained Weak Rank Principle. The polynomial method used in the proof of the SoS upper bound for the Norm-Constrained Weak Rank Principle stated in Theorem~\ref{thm:norm-constraints-upper-bound}, and the proof of the SoS lower bound for the Norm-Constrained Weak Rank Principle stated in Theorem~\ref{thm:norm-constraints-lower-bound} using a special pseudoexpectation operator independent of the norm budget were proposed by GPT-6 Astra Ultra.

\section{Main Results}

\subsection{Complexity of Weak Rank and Other Linear-Algebraic Principles}

Our first result determines the exact SoS degree complexity of the Weak Rank Principle.

\begin{theorem}\label{thm:WRank-thm}
    Let $n > r$ be positive integers and $\AA \in \bR^{n \times n}$ be a matrix of rank at least $r+1$. There exists a degree $2r+2$ SoS proof of $\WRank_r^n(\AA)$. Moreover, there does not exist an SoS proof of $\WRank_r^n(\AA)$ of degree at most $2r+1$.
\end{theorem}

Our next result does the same but with a constant-degree SoS proof for $\TFrame$, in contrast to the linear degree needed for $\WRank$. Recall that $\TFrame_r^n$ says that $\sX\sX^\top = \II_n$ is infeasible for $n \times r$ matrix $\sX$ when $n > r$, which only differs from $\WRank_r^n(\II_n)$ by changing the matrix $\sY$ to be the same as $\sX$ in the system $\sX\sY^\top = \II_n$.

\begin{theorem}\label{thm:TFrame-thm}
    Let $n > r \geq 1$ be integers. Then, there exists a degree-$4$ SoS proof of $\TFrame_r^n$.
\end{theorem}
\noindent
Because of the reduction in Proposition~\ref{prop:proof-system-reduction}, the above result also gives a degree-$4$ SoS refutation of $\WPHP_r^n$ when $n > r$.

We also obtain analogous SoS degree complexity results for the related inversion and orthogonality principles.

\begin{theorem}\label{thm:Inv-thm}
    Let $n \geq 1$ be an integer. Then, there exists a degree $2n+2$ SoS proof of $\Inv_n$, and there does not exist an SoS proof of $\Inv_n$ of degree at most $2n+1$.
\end{theorem}

\begin{theorem}\label{thm:Orth-thm}
    Let $n \geq 1$ be an integer. Then, there exists a degree $4$ SoS proof of $\Orthog_n$.
\end{theorem}

\subsection{Complexity of Weak Rank Principle with Norm Constraints}

Motivated by the drastic difference in the degree needed by an SoS refutation of the Weak Rank Principle $\sX\sY^\top = \II_n$ and that of the Tight Frame Principle $\sX\sX^\top = \II_n$, we next provide some preliminary investigation of why the real Weak Rank Principle is so hard for the SoS proof system.

Taking the trace of the Tight Frame equations $\sX\sX^\top=\II_n$ gives the Frobenius norm constraint $\|\sX\|_F^2=n$.
This is a degree-$2$ certificate of the \emph{Archimedean property}: for equations $p_1(\xx)=\cdots=p_s(\xx)=0$, this property means that $C-\|\xx\|_2^2=\sigma+\sum_{i=1}^s h_i p_i$ for some constant $C>0$, sum-of-squares polynomial $\sigma$, and polynomials $h_i$, certifying that every solution of a polynomial system lies in a ball of some radius.
In SoS relaxations for polynomial optimization, this property for instance appears as a hypothesis in Putinar's Positivstellensatz, which for systems of equations guarantees that every polynomial strictly positive on the feasible set is a sum of squares modulo the defining equations~\cite{putinar1993positive}.

The Weak Rank Principle equations $\sX\sY^\top=\II_n$, on the other hand, admit no degree-$2$ certificate of a bound on $\|\sX\|_F^2+\|\sY\|_F^2$.
We therefore propose to study how including an explicit norm bound affects the SoS refutation degree.
To do so, we introduce the following variant of the Weak Rank Principle.

\begin{definition}[Norm-Constrained Weak Rank Principle]
    Let $B \in \bR$. Let $\sX, \sY$ be two $n \times r$ matrices of real indeterminates, and $\st$ be a real indeterminate. The Norm-Constrained Weak Rank Principle, denoted $\NormWRank_{r}^{n,B}$ states that for an arbitrary matrix $\AA \in \bR^{n \times n}$ of rank at least $r + 1$, the system
    \begin{align*}
        \sX\sY^\top &= \AA,\\
        \|\sX\|_F^2 + \|\sY\|_F^2 + \st^2 &= B
    \end{align*}
    is infeasible. We use $\NormWRank_{r}^{n,B}(\AA)$ to denote the infeasible system parameterized by $\AA \in \bR^{n \times n}$ and $B \in \bR$.
\end{definition}

We show that, unlike the Weak Rank Principle which needs an SoS degree $\Omega(r)$ to prove, given the norm constraint $\|\sX\|_F^2 + \|\sY\|_F^2 \le B$, the Norm-Constrained Weak Rank Principle can have an SoS proof at a degree potentially much smaller than $r$, provided $B$ is small enough.

\begin{theorem}\label{thm:norm-constraints-upper-bound}
    Let $n > r \geq 1$ be integers, and $B \in \bR$ satisfy that $B > 2n$. Let $D \ge 1$ be an integer that satisfies
    \begin{align*}
        (n-r)\sinh^2\left((2D-1)\arcsinh\left(\frac{2}{\sqrt{B-2n}}\right)\right) > \frac{B-2n}{4},
    \end{align*}
    then there exists a degree-$4D$ SoS proof of $\NormWRank_r^{n,B}(\II_n)$.
\end{theorem}

Working out the dependency of $D$ on $B-2n$, we get the following quantitative bound. In particular, whenever $r = \omega\left(\sqrt{n}\log(n)\right)$ and $B = O(n)$, $\NormWRank_r^{n,B}(\II_n)$ has an SoS proof of degree $o(r)$, much smaller than the SoS degree need by $\WRank_r^n(\II_n)$.
\begin{corollary} \label{cor:norm-constraints-upper-bound}
    Fix a constant $C > 2$, and let $n > r \geq 1$ be integers. If $B \in \bR$ satisfy that $2n < B < Cn$, then there exists an SoS proof of $\NormWRank_r^{n,B}(\II_n)$ at degree $O_{C}(\sqrt{n}\log(n))$, where the implicit constant only depends on $C$.
\end{corollary}

\begin{remark}
    Note that if $B = 2n$, then the constraints $\sX\sY^\top = \II_n$ would force $\|\sX - \sY\|_F^2 = \|\sX\|_F^2 + \|\sY\|_F^2 - 2\tr(\sX\sY^\top) \equiv \|\sX\|_F^2 + \|\sY\|_F^2 - B \equiv -\st^2$ modulo the constraints. We would then be able to use the degree-$4$ SoS refutation for $\sX\sX^\top = \II_n$ to refute the feasibility of polynomial system in the Norm-Constrained Weak Rank Principle when $B=2n$.
\end{remark}

However, adding a norm constraint does not always lead to an SoS proof of lower degree. We prove that, on the other hand, if the norm bound $B$ is large enough, then the polynomial system of $\NormWRank_r^{n,B}(\II_n)$ still needs SoS degree exactly $2r+2$ to refute, matching the complexity of $\WRank_{r}^n(\II_n)$.

\begin{theorem}\label{thm:norm-constraints-lower-bound}
    Let $n > r$ be positive integers. There exists $B = B(n,r) \in \bR$ large enough such that there does not exist an SoS proof of $\NormWRank_r^{n,B}(\II_n)$ of degree at most $2r+1$.
\end{theorem}

This quantitative dependence of proof complexity on the radius of an Archimedean property has been proposed and observed numerically in some prior work \cite{henrion2005detecting,schweighofer2006global}.
However, we are not aware of previous work that has rigorously established such a quantitative tradeoff, and we believe this is an interesting direction for future investigation.

\subsection{Applications}

Lastly, we also provide SoS lower bounds for the following problems as applications of our SoS lower bound for the Weak Rank Principle.

\subsubsection{Sign Rank}
\begin{definition}[Sign Rank]
    For a sign matrix $\SS \in \{\pm 1\}^{n \times n}$, its sign rank
    \[\signrank(\SS) \colonequals \min\{r \ge 0: \exists \AA \in \bR^{n \times n}, \rank(\AA) = r, \text{ and } \SS_{i,j} = \sign(\AA_{i,j}) \forall\, 1\le i,j\le n\}\]
    is the smallest integer $r \ge 0$ such that there exists a rank-$r$ matrix $\AA \in \bR^{n \times n}$ whose entry-wise sign agrees with $\SS$.
\end{definition}
\noindent
We consider the following polynomial encoding of the sign rank of a matrix being at most some $r$.

\begin{definition}[Polynomial Encoding of Sign Rank]
    Let $\SS \in \{\pm 1\}^{n \times n}$ be a sign matrix. Let $\sX, \sY$ be two $n \times r$ matrices of real indeterminates, and let $\sZ$ be an $n \times n$ matrices of real indeterminates. $\SRank_r^n(\SS)$ is the following system of polynomials:
    \begin{align*}
        \SS_{i,j} \sum_{k=1}^r \sX_{i,k}\sY_{j,k} = 1 + \sZ_{i,j}^2, && \forall\, (i,j) \in [n]^2.
    \end{align*}
\end{definition}

Clearly, whenever $\SRank_r^n(\SS)$ has a real solution, then $\signrank(\SS) \le r$. Conversely, we note that if $\signrank(\SS) \le r$, there exists a matrix that can be factorized as $\XX\YY^\top$ where $\XX,\YY \in \bR^{n \times r}$ such that the entry-wise product of $\SS$ and $\XX\YY^\top$ is at least $1$.

\begin{corollary}\label{cor:sign-rank}
    Let $r$ be a positive integer and $\SS \in \{\pm 1\}^{n \times n}$ be a matrix of sign rank at least $r+1$. Then, $\SRank_r^n(\SS)$ does not have an SoS refutation of degree at most $2r+1$.
\end{corollary}

\subsubsection{Matrix Rigidity}

\begin{definition}[Matrix Rigidity]
    For $\AA \in \bR^{n \times n}$ and an integer $r \ge 0$, its rigidity
    \begin{align*}
        \calR_{\AA}(r) \colonequals \min\{\#\supp(\EE): \EE \in \bR^{n \times n}, \rank(\AA - \EE) \le r \}
    \end{align*}
    is defined as the minimum number of entry changes that turn $\AA$ into a matrix of rank at most $r$.
\end{definition}
\noindent
We consider the following polynomial encoding of the rigidity of a matrix being at most some $s$ (for a given $r$).

\begin{definition}[Polynomial Encoding of Rigidity]
    Let $s \ge 0$ and $r \ge 1$ be integers and $\AA \in \bR^{n \times n}$ be a matrix. Let $\sX,\sY$ be two $n \times r$ matrices of real indeterminates, let $\sE, \sZ$ be two $n \times n$ matrix of real indeterminates, and let $\st$ be a real indeterminate. $\Rigid_r^s(\AA)$ is then the following system of polynomials:
    \begin{align*}
        \sX\sY^\top + \sE &= \AA,\\
        \sE_{i,j}(1 - \sZ_{i,j}) &= 0, && \forall\, (i,j) \in [n]^2\\
        \sZ_{i,j}^2 - \sZ_{i,j} &= 0, && \forall\, (i,j) \in [n]^2\\
        \sum_{i,j=1}^n \sZ_{i,j} &= s - \st^2.
    \end{align*}
\end{definition}

In the encoding above, $\sE$ encodes the difference between $\AA$ and some rank-$r$ matrix, $\sZ$ encodes the support of $\sE$, and $\sum_{i,j=1}^n \sZ_{i,j} = s - \st^2$ encodes that the size of the support of $\sZ$ is at most $s$.

\begin{corollary}\label{cor:matrix-rigidity}
    Let $s \ge 0$ and $r \ge 1$ be integers and $\AA \in \bR^{n \times n}$ be a matrix such that $\calR_{\AA}(r) > s$. Then, $\Rigid_r^s(\AA)$ does not have an SoS refutation of degree at most $2r+1$.
\end{corollary}

\subsubsection{Algebraic Branching Programs}

\begin{definition}[Non-Commutative Algebraic Branching Program]
    We call a homogeneous non-commutative algebraic branching program (ABP) over $\mathbb{R}$ a directed acyclic graph with a source and sink, whose vertices are partitioned into layers $V_0, \dots, V_{\ell}$ where directed edges always go from some layer $V_{i-1}$ to $V_i$. $V_0$ contains a single vertex designated as the source $s$, and $V_{\ell}$ contains a single vertex designated as the sink $t$. 
    The width of an ABP is the largest size of its layers $V_i$ for $0 \le i \le \ell$.

    Every edge $e$ of the directed graph is further labelled with a linear form $L_e(\zz) = \sum_{j=1}^n a_{e,j} z_j$, where $a_{e,j} \in \bR$ and $z_j$ are non-commutative formal variables. The ABP computes a non-commutative polynomial $f \in \bR\langle z_1, \dots, z_n \rangle$ given by
    \[f = \sum_{\substack{\text{path } P\\ \text{from } s \text{ to } t}} \prod_{e \in P} L_e(\zz),\]
    where the product is taken according to the path order.
    Equivalently, $f$ can be computed recursively as follows:
    \begin{itemize}
        \item At the start, assign the constant polynomial $f_s = 1$ to the source $s \in V_0$.
        \item For $1 \le i \le \ell$, for every vertex $v \in V_i$, compute
        \begin{align*}
            f_v = \sum_{\substack{\text{edge }e = (u \to v):\\ u \in V_{i-1} }} f_{u} L_e(\zz).
        \end{align*}
        \item Output the computed polynomial $f_t$ at the sink.
    \end{itemize}
\end{definition}

We follow the recursive encoding of \cite[Definition 8.6]{garlik2026weak} and consider the following polynomial system for the existence of a homogeneous non-commutative ABP over the reals of width $r$ that computes a given non-commutative polynomial $f \in \bR\langle z_1, \dots, z_n\rangle$.

\begin{definition}[Polynomial Encoding of Existence of Small-Width ABP]
    Let $r \ge 1$ be an integer, and $f \in \bR\langle z_1, \dots, z_n\rangle$ be a homogeneous non-commutative polynomial given by
    \begin{align*}
        f = \sum_{w: |w| = \ell } c_w \cdot w,
    \end{align*}
    where the summation is over all length-$\ell$ words $w$ in the variables $z_1, \dots, z_n$, for some coefficients $c_w \in \bR$. Denote $(r_0, r_1, \dots, r_{\ell-1}, r_{\ell}) \colonequals (1, r, \dots, r, 1)$. For every $1\le i \le \ell$ and $1 \le j \le n$, let $\sE^{(i,z_j)}$ be an $r_{i-1} \times r_i$ matrix of real indeterminates. For every $1\le i \le \ell$ and word $w$ of length $i$, let $\sP^{(i,w)}$ be a length-$r_i$ row vector of real indeterminates. Then, $\ncABP(f,r)$ is the following system of polynomials:
    \begin{align*}
        \sP^{(1,z_j)} &= \sE^{(1,z_j)}, && \forall\, j \in [n]\\
        \sP^{(i,wz_j)}  &= \sP^{(i-1,w)}\sE^{(i,z_j)}, && \forall\, 2\le i \le \ell, j\in [n], |w| = i-1\\
        \sP^{(\ell,w)} &= c_w, && \forall\, |w| = \ell.
    \end{align*}
\end{definition}

To see that $\ncABP(f,r)$ is a polynomial encoding of the existence of a non-commutative ABP over the reals of width at most $r$ that computes $f$, we note that the computation of the ABP can be modelled as a product of matrices whose entries specify the linear forms on the directed edges from one layer to the next.
That is, the non-commutative polynomial the ABP computes is
\begin{align*}
    M^{(1)}(\zz) M^{(2)}(\zz) \dots M^{(\ell)}(\zz),
\end{align*}
where $M^{(i)}(\zz)$ is an $r_{i-1} \times r_i$ matrix, whose $(u,v)$-entry is the linear form on the edge $u \to v$. We may then take $\sE^{(i,z_j)}$ to be the encoding of the coefficient matrix of $z_j$ in $M^{(i)}(\zz)$ so that $M^{(i)}(\zz) = \sum_{j=1}^n \sE^{(i,z_j)} z_j$. Under this encoding, we see that $\sP^{(i,w)}$ encodes the coefficient row vector of the word $w$ at each vertex in the layer $V_i$, and the last constraint $\sP^{(\ell,w)} = c_{w}$ asks that the coefficients of the computed polynomial at the sink $t$ match the coefficients of $f$. 

\begin{corollary}\label{cor:ABP}
    Let $r \geq 1$ be an integer and $f \in \bR\langle z_1, \dots, z_n\rangle$ be a homogeneous non-commutative polynomial of degree $\ell$. Then, $\ncABP(f,r)$ does not have an SoS refutation of degree at most $r+1$.
\end{corollary}

\subsubsection{Tensor CP Rank}

\begin{definition}[Tensor CP Rank]
    Let $d \ge 2$ and $\TT \in \bR^{n_1 \times \dots \times n_d}$ be a $d$-tensor. Its real CP rank is the smallest integer $r \ge 0$ such that $\TT$ can be written as a sum of $r$ tensor products of real vectors, i.e.,
    \begin{align*}
        \rank_{\CP}(\TT) \colonequals \min\left\{r \ge 0: \exists\, \vv_i^{(t)} \in \bR^{n_t} \text{ for } 1\le i \le r, 1\le t \le d, \text{ s.t. } \TT = \sum_{i=1}^r \vv_i^{(1)} \otimes \dots \otimes \vv_i^{(d)}\right\}.
    \end{align*}
\end{definition}
\noindent
We consider the following polynomial encoding of the tensor CP rank being at most some $r$.
\begin{definition}[Polynomial Encoding of Tensor CP Rank]
    Let $d \ge 2$, $r \ge 1$ be integers and $\TT \in \bR^{n_1 \times \dots \times n_d}$ be a $d$-tensor. For $1 \le t \le d$, let $\sV^{(t)}$ be $n_t \times r$ matrices of real indeterminates. $\CPRank_r^{(n_1, \dots, n_d)}(\TT)$ is the following system of polynomials:
    \begin{align*}
        \hspace{0.4in}\sum_{i=1}^r \prod_{t=1}^d \sV^{(t)}_{a_t,i} = \TT_{a_1, \dots, a_d}, && \forall\, (a_1, \dots, a_d) \in [n_1] \times \dots \times [n_d].
    \end{align*}
\end{definition}

\begin{corollary}\label{cor:CPRank}
    Let $d \ge 3, r \ge 1$ be integers and $\TT \in \bR^{n_1 \times \dots \times n_d}$ be a $d$-tensor with real CP rank at least $r+1$. Suppose that $n_1 \ge n_2 \ge \dots \ge n_d$. Then, $\CPRank_r^{(n_1, \dots, n_d)}(\TT)$ does not have an SoS refutation of degree at most $$\max_{2\le \ell \le d} \left\lfloor \left(\frac{r}{N_{\ell}}\right)^{\frac{1}{\ell-1}}\right\rfloor + (d - \ell) + 1,$$
    where $N_{\ell} \colonequals \prod_{j = \ell+1}^d n_j$.
\end{corollary}

\subsection{Proof Overview of Theorem~\ref{thm:WRank-thm}: Complexity of Weak Rank Principle}

The proof of our SoS upper bound for $\WRank_r^n$ follows from a straightforward determinantal refutation, which provides a degree $2r+2$ Nullstellensatz refutation. See Section~\ref{sec:WRank-upper-bound} for details. Now we explain the proof idea behind the SoS lower bound for $\WRank_r^n$.

By the reduction from $\WRank_r^n(\AA)$ to $\WRank_r^n(\II_n)$ given in Proposition~\ref{prop:proof-system-reduction}, it suffices to prove the SoS lower bound for $\WRank_r^n(\II_n)$. Our SoS lower bound proof proceeds by showing a contradiction \emph{without} constructing an explicit degree-$(2r+1)$ pseudoexpectation operator.

Given a hypothetical SoS refutation $-1 = g + \sum_{1\le i,j\le n} h_{i,j} \cdot ((\sX\sY^\top)_{i,j} - \delta_{i,j})$ of degree at most $2r+1$ where $g$ is a sum-of-squares polynomials, taking the average of the refutation at $(\sX,\sY)$ and at $(-\sX,-\sY)$ cancels out the odd-degree components of $h_{i,j}$, giving an SoS refutation of $\sX\sY^\top = \II_n$ of degree at most $2r$.

The strategy of the proof is then as follows. Given the polynomial identity $-1 = g + \sum_{1\le i,j\le n} h_{i,j} \cdot ((\sX\sY^\top)_{i,j} - \delta_{i,j})$ where every polynomial on the right hand side has degree at most $2r$, we show that the sum-of-squares polynomial $g$ cannot be a nonnegative constant. We establish this using Proposition~\ref{prop:vanishing-top-degree}, which says that if $-1-g$ is a constant polynomial that can be expressed by the constraints $\sX\sY^\top = \II_n$ at degree $2r$, then $-1-g$ vanishes at the evaluation point $(\zzero, \zzero)$. This would lead to a contradiction.

Thus, the sum-of-squares polynomial $g$ in the SoS refutation must have degree $2d$ for some $1\le d\le r$. We choose an extremal degree-$2r$ SoS refutation of $\sX\sY^\top = \II_n$ where the sum-of-squares polynomial $g$ has the smallest degree $2d$. We will show that we can always find another degree-$2r$ SoS refutation of $\sX\sY^\top = \II_n$ in which the sum-of-squares polynomial has a smaller degree, contradicting the extremality of $g$. The degree reduction step goes as follows. Again we write $-1 - g = \sum_{1\le i,j\le n} h_{i,j} \cdot ((\sX\sY^\top)_{i,j} - \delta_{i,j})$, so that the degree-$2d$ polynomial $-1 - g$ can be expressed by the constraints $\sX\sY^\top = \II_n$ at degree $2r$. Proposition~\ref{prop:vanishing-top-degree} now implies that the degree-$2d$ component of $-g$ vanishes on the set of real matrix pairs $(\XX,\YY) \in \bR^{n \times r} \times \bR^{n \times r}$ satisfying $\XX\YY^\top = \zzero$ and $\rank(\XX) + \rank(\YY) \le d$. In particular, Proposition~\ref{prop:vanishing-structure} says that the degree-$d$ component of every polynomial $q_{\alpha}$ in $g = \sum_{\alpha} q_\alpha^2$ can be written as a linear combination of $(\sX\sY^\top)_{i,j}$. Using this fact, we peel off the degree-$d$ component of each $q_{\alpha}$ to reduce the degree of the sum-of-squares polynomial $g$ while preserving an SoS refutation at degree $2r$. This degree reduction step leads to a contradiction, proving that no such SoS refutation of $\sX\sY^\top = \II_n$ exists at degree $2r$.

The main technical observations are Proposition~\ref{prop:vanishing-top-degree} and~\ref{prop:vanishing-structure}, which can be viewed as structural results about polynomials on $\mathcal{Z}_d$.
Proposition~\ref{prop:vanishing-top-degree} is obtained by a construction of an explicit pseudoexpectation-like object associated to the Nullstellensatz, without a positivity constraint.
Proposition~\ref{prop:vanishing-structure}, which is perhaps the key step of the proof, shows that low-degree polynomials vanishing on $\mathcal{Z}_d$ are the same as low-degree polynomials vanishing on all of $\{\bm X\bm Y^{\top} = \bm 0\}$ with no extra rank constraint.
Its proof decomposes a polynomial by degree in each row, so each component involves at most $d$ rows, and applies a result of~\cite{CW19} establishing primality of the corresponding bilinear ideal.
A rational parametrization expresses each component using the bilinear generators with powers of a determinant in the denominators, and primality then gives membership in the ideal with polynomial multipliers.

\section{Notations}

Let $f \in \bR[\sX]$ be a multivariate polynomial. We use the convention that the $0$ polynomial has degree $-\infty$, so that if $\deg(f) \le d_1$ and $\deg(g) \le d_2$, then $\deg(f\cdot g) \le d_1 + d_2$ and $\deg(f + g) \le \max\{d_1, d_2\}$. For a nonnegative integer $d$, we use $f^{[d]} \in \bR[\sX]$ to denote the homogeneous component of $f$ of degree $d$. If $\deg(f) \le D$, then
\begin{align*}
    f(\sX) = \sum_{d=0}^D f^{[d]}(\sX).
\end{align*}

For $n, r, d \geq 1$ we denote the rank-bounded zero locus $\calZ_d$ as
\begin{equation}
    \calZ_d \colonequals \{(\XX,\YY) \in \bR^{n \times r} \times \bR^{n \times r}: \XX\YY^\top = \zzero,  \rank(\XX) + \rank(\YY) \le d\}. \label{eq:Zd}
\end{equation}
For $n \times r$ matrix indeterminates $\sX, \sY$, we say a polynomial $f \in \bR[\sX, \sY]$ vanishes on $\calZ_d$ if for any $(\XX, \YY) \in \calZ_d$ we have $f(\XX, \YY) = 0$.

\section{SoS Complexity of $\WRank$: Proof of Theorem~\ref{thm:WRank-thm}}

\subsection{Upper Bound} \label{sec:WRank-upper-bound}

At degree $2r + 2$, an SoS proof for $\WRank_r^n(\AA)$ exists by considering the determinant of an $(r+1) \times (r+1)$ invertible submatrix of $\AA$.
In fact, as we will see we may construct a Nullstellensatz such proof, i.e., an SoS proof without sum-of-squares terms.

\begin{proof}
    Let $k=r+1$ and choose an invertible submatrix $\BB=\AA_{I,J}$ of size $k\times k$.
    Set $\sM=\sX_{I,[r]}\sY_{J,[r]}^\top$.
    The equations $\sM_{i,j}-\BB_{i,j}=0$ for $i,j\in[k]$ form a subsystem of $\sX\sY^\top=\AA$.
    Since $\sM$ is a product of a $k\times r$ matrix and an $r\times k$ matrix, $\det(\sM)=0$ identically.

    For $0\le j\le k$, let $\sM^{(j)}$ have its first $j$ columns from $\sM$ and its remaining columns from $\BB$.
    Then $\sM^{(0)}=\BB$ and $\sM^{(k)}=\sM$.
    For $1\le j\le k$, the matrices $\sM^{(j)}$ and $\sM^{(j-1)}$ differ only in column $j$, so expanding their determinants along this column gives
    \[
        \det(\sM^{(j)})-\det(\sM^{(j-1)})
        =\sum_{i=1}^k H_{i,j}(\sM_{i,j}-\BB_{i,j}),
    \]
    where $H_{i,j}$ is the $(i,j)$ cofactor of $\sM^{(j-1)}$.
    Each cofactor is, up to sign, a determinant of size $r$ whose entries are constants or quadratic polynomials in $\sX,\sY$.
    Thus $\deg(H_{i,j})\le2r$.

    Summing over $j$ and using $\det(\sM)=0$ gives
    \[
        -\det(\BB)=\sum_{i,j=1}^k H_{i,j}(\sM_{i,j}-\BB_{i,j}).
    \]
    Since $\det(\BB)\ne0$, we may divide to obtain
    \[
        -1=\sum_{i,j=1}^k\frac{H_{i,j}}{\det(\BB)}(\sM_{i,j}-\BB_{i,j}).
    \]
    Each summand has degree at most $2r+2$.
    This is a Nullstellensatz refutation of the subsystem, and hence of $\sX\sY^\top=\AA$.
    Taking the sum-of-squares term to be zero gives the required degree-$(2r+2)$ SoS refutation.
\end{proof}

\subsection{Auxiliary Results}

Our lower bound builds on the following two propositions.

\begin{proposition} \label{prop:vanishing-top-degree}
    Let $d \le r < n$, where $n$ and $r$ are positive integers and $d$ is a nonnegative integer. Let $\sX, \sY$ be $n \times r$ matrices of real indeterminates, and $f \in \bR[\sX, \sY]$ be a polynomial of degree at most $2d$. If there exist polynomials $h_{i,j} \in \bR[\sX, \sY]$ for $1 \le i,j\le n$ of degrees at most $2r-2$ such that
    \begin{align*}
        f = \sum_{1\le i,j\le n} h_{i,j} \cdot \left((\sX \sY^\top)_{i,j} - \delta_{i,j}\right),
    \end{align*}
    then $f^{[2d]}$ vanishes on $\calZ_{d}$.
\end{proposition}

\begin{proof}
    We first construct a linear functional that rules out low-degree representations of nonzero constants.
    This object may be viewed as a dual witness ruling out low-degree Nullstellensatz refutations.
    For a positive integer $q$, let $\sU,\sV$ be $n \times q$ matrices of indeterminates.
    Define $L_q:\bR[\sU,\sV]_{\le 2q}\to\bR$ on monomials by
    \[
        L_q\left[\prod_{a\in S}\sU_{i_a,a}\sV_{j_a,a}\right]
        = \frac{(q-|S|)!}{q!}\prod_{a\in S}\delta_{i_a,j_a}
    \]
    for $S\subseteq[q]$ and $i_a,j_a\in[n]$, set it to zero on every other monomial, and extend $L_q$ linearly.
    On the empty product we set $L_q[1]=1$.

    We claim that
    \[
        L_q\left[h \cdot (\sU\sV^\top)_{i,j}\right]=\delta_{i,j}L_q[h]
    \]
    whenever $\deg(h)\le 2q-2$.
    By linearity, it suffices to check this for a monomial $h$.
    If $h=\prod_{a\in S}\sU_{i_a,a}\sV_{j_a,a}$ and $k=|S|$, then $k\le q-1$.
    In the expansion of $h \cdot (\sU\sV^\top)_{i,j}$, the terms that repeat a column index receive value zero.
    Each of the $q-k$ remaining terms receives value
    \[
        \frac{(q-k-1)!}{q!}\delta_{i,j}\prod_{a\in S}\delta_{i_a,j_a},
    \]
    so their sum is $\delta_{i,j}L_q[h]$.
    For any other monomial $h$, none of the products $h \cdot \sU_{i,a}\sV_{j,a}$ has this form, and so both sides are zero.

    It follows that any constant $c$ satisfying
    \[
        c=\sum_{i,j=1}^n g_{i,j} \cdot \left((\sU\sV^\top)_{i,j}-\delta_{i,j}\right)
    \]
    with $\deg(g_{i,j})\le 2q-2$ must be zero, by applying the functional $L_q$ to either side of the identity.

    If $d=0$, applying this fact with $q=r$ gives $f=0$.
    If $d=r$, taking the degree-$2r$ component of the assumed identity gives
    \[
        f^{[2r]}=\sum_{i,j=1}^n h_{i,j}^{[2r-2]}(\sX\sY^\top)_{i,j},
    \]
    which vanishes whenever $\XX\YY^\top=\zzero$.
    We may therefore assume $0<d<r$.

    Now, fix $(\XX,\YY)\in\calZ_d$ and write $a=\rank(\XX)$, $b=\rank(\YY)$, and $q=r-a-b$.
    Then $q\ge r-d\ge1$.
    The row spaces of $\XX$ and $\YY$ are mutually orthogonal, since $\XX\YY^\top=\zzero$.
    Choose an orthogonal matrix $\OO$ whose columns are orthonormal bases of these two row spaces, followed by a basis of their common orthogonal complement.
    Thus
    \begin{align*}
        \XX\OO&=\begin{pmatrix}\XX_1&\zzero&\zzero\end{pmatrix},\\
        \YY\OO&=\begin{pmatrix}\zzero&\YY_2&\zzero\end{pmatrix},
    \end{align*}
    where the column blocks have sizes $a,b,q$.
    Let $\sX_1,\sY_2,\sX_3,\sY_3$ be matrices of independent indeterminates with the corresponding sizes.
    Define
    \[
        \widetilde f(\sX_1,\sY_2,\sX_3,\sY_3)
        \colonequals f\left(
        \begin{pmatrix}\sX_1&\zzero&\sX_3\end{pmatrix}\OO^\top,
        \begin{pmatrix}\zzero&\sY_2&\sY_3\end{pmatrix}\OO^\top\right),
    \]
    and define $\widetilde h_{i,j}$ by the same substitution in $h_{i,j}$.
    The substitution does not increase degrees, so $\deg(\widetilde f)\le2d$ and $\deg(\widetilde h_{i,j})\le2r-2$.
    The identity for $f$ becomes
    \[
        \widetilde f=\sum_{i,j=1}^n\widetilde h_{i,j}\cdot
        \left((\sX_3\sY_3^\top)_{i,j}-\delta_{i,j}\right),
    \]
    whose equation factors depend only on the last blocks.

    Let $\mu$ be a monomial of degree $2d$ in the entries of $\sX_1,\sY_2$.
    Its coefficient in $\widetilde f$ is a constant $c_\mu$, since $\deg(\widetilde f)\le2d$.
    Let $g_{i,j,\mu}$ be its coefficient in $\widetilde h_{i,j}$, viewed as a polynomial in $\sX_3,\sY_3$.
    These coefficients satisfy
    \[
        \deg(g_{i,j,\mu})\le2r-2-2d=2(r-d)-2\le2q-2.
    \]
    Taking the coefficient of $\mu$ in the identity gives
    \[
        c_\mu=\sum_{i,j=1}^n g_{i,j,\mu} \cdot
        \left((\sX_3\sY_3^\top)_{i,j}-\delta_{i,j}\right).
    \]
    Applying $L_q$ to the last blocks gives $c_\mu=0$.

    Thus every degree-$2d$ term of $\widetilde f$ contains a variable from $\sX_3$ or $\sY_3$, so
    \[
        \widetilde f^{[2d]}(\sX_1,\sY_2,\zzero,\zzero)=0.
    \]
    Since the substitution is linear, it commutes with taking homogeneous components.
    Evaluating the remaining blocks at $\XX_1,\YY_2$ therefore gives
    \[
        f^{[2d]}(\XX,\YY)
        =\widetilde f^{[2d]}(\XX_1,\YY_2,\zzero,\zzero)=0,
    \]
    as required.
\end{proof}

\begin{proposition} \label{prop:vanishing-structure}
    Let $d \le r < n$, where $n$ and $r$ are positive integers and $d$ is a nonnegative integer. Let $\sX, \sY$ be $n \times r$ matrices of real indeterminates, and $f \in \bR[\sX, \sY]$ be a polynomial of degree at most $d$. If $f$ vanishes on $\calZ_d$, then there exist polynomials $h_{i,j} \in \bR[\sX, \sY]$ for $1 \le i,j \le n$ such that
    \begin{align*}
        f = \sum_{1\le i, j \le n} h_{i,j} \cdot (\sX\sY^\top)_{i,j},
    \end{align*}
    and $\deg(h_{i,j}) \le d-2$ for all $1\le i, j \le n$.
\end{proposition}

\begin{proof}
    Group the monomials of $f$ according to their degrees in each row of $\sX$ and $\sY$.
    Each resulting component is homogeneous in every row separately.
    Scaling the rows independently preserves $\calZ_d$.
    Thus, at any point of $\calZ_d$, evaluating $f$ after these scalings gives a polynomial in the scaling factors that vanishes identically.
    Its coefficients are the values of the components, so each component also vanishes on $\calZ_d$.

    Fix a nonzero component $F$, and let $I,J\subseteq[n]$ be the rows of $\sX,\sY$, respectively, in which $F$ has positive degree.
    Every such row contributes at least one to the total degree, so
    \[
        |I|+|J|\le\deg(F)\le d\le r.
    \]
    Any real assignment of these rows satisfying $(\XX\YY^\top)_{i,j}=0$ for $i\in I,j\in J$ extends to a point of $\calZ_d$ by setting all other rows to zero.
    Indeed, the extended matrices have zero product and their ranks sum to at most $|I|+|J|$.
    Hence $F$ vanishes on every real solution of these equations.
    Both $I$ and $J$ must be nonempty, since otherwise every assignment would be allowed and $F$ would be zero.

    Consider the ideal
    \[
        \mathfrak a\colonequals
        \left\langle (\sX\sY^\top)_{i,j}:i\in I,j\in J\right\rangle
        \subseteq\bR[\sX_{I,[r]},\sY_{J,[r]}].
    \]
    Since $|I|+|J|\le r$, this ideal is prime by \cite[Theorem~8.6(3)]{CW19}, applied to the complete bipartite graph $K_{|I|,|J|}$.
    To show that $F\in\mathfrak a$, we first express $F$ as a combination of the generators allowing denominators that are powers of a fixed determinant $\Delta$.
    Clearing denominators will give $\Delta^N F\in\mathfrak a$, after which primality will allow us to remove the determinant factor.

    Choose the first $|I|$ columns, $C=\{1,\ldots,|I|\}\subseteq[r]$, and let $C^c=[r]\setminus C$ and $\Delta=\det(\sX_{I,C})$.
    For each $j\in J$, the equations $(\sX\sY^\top)_{I,j}=\zzero$ read
    \[
        \sX_{I,C}\sY_{j,C}^\top+\sX_{I,C^c}\sY_{j,C^c}^\top=\zzero.
    \]
    When $\Delta\ne0$, these determine $\sY_{J,C}$ uniquely from $\sX_{I,[r]}$ and $\sY_{J,C^c}$.
    Define the resulting matrix of rational functions by
    \begin{align*}
        \widetilde{\sY}_{J,C}^\top
        &\colonequals-\sX_{I,C}^{-1}\sX_{I,C^c}\sY_{J,C^c}^\top,\\
        \widetilde{\sY}_{J,C^c}&\colonequals\sY_{J,C^c}.
    \end{align*}
    For any real choice of the free entries $\sX_{I,[r]}$ and $\sY_{J,C^c}$ with $\Delta\ne0$, the pair $(\sX_{I,[r]},\widetilde{\sY}_{J,[r]})$ solves all the equations.

    Regard $F$ as a polynomial in the rows indexed by $I,J$, and define
    \[
        \widetilde F(\sX_{I,[r]},\sY_{J,C^c})
        \colonequals F(\sX_{I,[r]},\widetilde{\sY}_{J,[r]}).
    \]
    Set $m=\deg(F)$ and define a denominator $Q$ and numerator $P$ by
    \begin{align*}
        Q&\colonequals\Delta^m,\\
        P&\colonequals Q\widetilde F,
    \end{align*}
    so that $\widetilde F=P/Q$.
    Each entry of $\sX_{I,C}^{-1}$ has denominator $\Delta$, and each monomial of $F$ contains at most $m$ substituted entries.
    Hence $P$ is a polynomial in the free variables.
    Since $F$ vanishes at every real solution, $P=Q\widetilde F$ vanishes for every real choice of these variables with $\Delta\ne0$.
    This is a nonempty open set, so $P=0$ identically and therefore $\widetilde F=0$.

    The differences between the original variables and their substituted values are the entries of
    \[
        \sY_{j,C}^\top-\widetilde{\sY}_{j,C}^\top
        =\sX_{I,C}^{-1}(\sX\sY^\top)_{I,j}.
    \]
    Replacing a variable $z$ by an expression $w$ changes a polynomial by a multiple of $z-w$.
    Replacing the variables one at a time therefore expresses $F-\widetilde F=F-P/Q$ as a combination of these differences, and hence of the generators, with denominators that are powers of $\Delta$.
    Choose a common denominator $\Delta^N$ with $N\ge m$.
    Multiplying by it and using $Q=\Delta^m$ gives
    \[
        \Delta^N F-\Delta^{N-m}P\in\mathfrak a.
    \]
    Since $P=0$, we obtain $\Delta^N F\in\mathfrak a$.
    All generators of $\mathfrak a$ vanish when $\sX_{I,C}=\II_{|I|}$ and all other entries are zero, whereas $\Delta=1$.
    Thus $\Delta\notin\mathfrak a$, and primality gives $F\in\mathfrak a$.

    Adding the representations for all components gives
    \[
        f=\sum_{i,j=1}^n h_{i,j}\cdot(\sX\sY^\top)_{i,j}
    \]
    for polynomial multipliers $h_{i,j}$.
    Each generator is homogeneous of degree two.
    Taking the components of degree at most $d$ in this identity therefore amounts to discarding all terms of degree greater than $d-2$ from each $h_{i,j}$.
    The left side remains $f$, giving the required degree bounds.
    For $d=0,1$, this shows that $f=0$, and all multipliers can be taken to be zero.
\end{proof}

\subsection{Lower Bound}

We now prove the lower bound side of Theorem~\ref{thm:WRank-thm}.
By Proposition~\ref{prop:proof-system-reduction}, it suffices to prove the lower bound for $\WRank_r^n(\II_n)$.

\begin{proof}
    Let $n > r$ be positive integers and fix a nonnegative integer $D\le2r+1$.
    Suppose for the sake of contradiction that $\WRank_r^n(\II_n)$ has a degree-$D$ SoS proof.

    Let $d$ be the smallest nonnegative integer for which such a proof has a sum-of-squares term of degree at most $2d$.
    Then $d\le\lfloor D/2\rfloor\le r$.
    Choose such a proof:
    \begin{align*}
        -1 &= g + \sum_{1\le i,j\le n} p_{i,j}\cdot \left((\sX\sY^\top)_{i,j} - \delta_{i,j}\right)\\
        &= \sum_{\alpha = 1}^m s_{\alpha}^2 + \sum_{1\le i,j\le n} p_{i,j}\cdot \left((\sX\sY^\top)_{i,j} - \delta_{i,j}\right),
    \end{align*}
    where $g \colonequals \sum_{\alpha = 1}^m s_{\alpha}^2 \in \bR[\sX,\sY]$, $\deg(s_{\alpha}) \le d$ for every $1\le \alpha \le m$, and $\deg(p_{i,j}) \le D-2$ for every $1\le i,j\le n$.
    Averaging this identity at $(\sX,\sY)$ and $(-\sX,-\sY)$ gives
    \begin{align*}
        -1 &= \sum_{\alpha = 1}^{2m} q_{\alpha}^2 + \sum_{1\le i,j\le n} h_{i,j}\cdot \left((\sX\sY^\top)_{i,j} - \delta_{i,j}\right), \numberthis \label{eq:min-deg-SoS-WRank}
    \end{align*}
    where we denote
    \begin{align*}
        q_{\alpha}(\sX, \sY) \colonequals \begin{cases}
            \frac{s_{\alpha}(\sX, \sY)}{\sqrt{2}}, &\quad \text{ for } 1 \le \alpha \le m,\\
            \frac{s_{\alpha-m}(-\sX, -\sY)}{\sqrt{2}}, &\quad \text{ for } m+1 \le \alpha \le 2m,
        \end{cases}
    \end{align*}
    and
    \begin{align*}
        h_{i,j}(\sX, \sY) \colonequals \frac{p_{i,j}(\sX, \sY) + p_{i,j}(-\sX, -\sY)}{2}, \quad \text{ for } 1 \le i,j \le n.
    \end{align*}
    The odd-degree components of each $p_{i,j}$ cancel in this average.
    Since $D\le2r+1$, this gives $\deg(h_{i,j})\le\min\{2r-2,D-2\}$ for every $i,j$.
    Thus \eqref{eq:min-deg-SoS-WRank} remains a degree-$D$ SoS proof, with $\deg(q_{\alpha})\le d$ for every $\alpha$.

    If $d=0$, then $\sum_{\alpha=1}^{2m}q_{\alpha}^2=C$ is a nonnegative constant.
    Rearranging and dividing by $1+C$ gives a Nullstellensatz refutation:
    \begin{align*}
        -1 &= \sum_{1\le i,j\le n} \frac{h_{i,j}}{1+C}\cdot \left((\sX\sY^\top)_{i,j} - \delta_{i,j}\right).
    \end{align*}
    Each multiplier $h_{i,j}/(1+C)$ has degree at most $2r-2$.
    Proposition~\ref{prop:vanishing-top-degree} would therefore imply that $(-1)^{[0]}=-1$ vanishes on $\calZ_0=\{(\zzero,\zzero)\}$, a contradiction.

    We may therefore assume that $d\ge1$.
    Set $f=1+\sum_{\alpha=1}^{2m}q_{\alpha}^2$.
    Equation~\eqref{eq:min-deg-SoS-WRank} gives
    \begin{align*}
        f = - \sum_{1\le i,j\le n} h_{i,j}\cdot \left((\sX\sY^\top)_{i,j} - \delta_{i,j}\right).
    \end{align*}
    We have $\deg(f)\le2d$ and $\deg(h_{i,j})\le2r-2$, so Proposition~\ref{prop:vanishing-top-degree} implies that $f^{[2d]}$ vanishes on $\calZ_d$.
    Since $d\ge1$, we have $f^{[2d]}=\sum_{\alpha=1}^{2m}(q_{\alpha}^{[d]})^2$.
    These are squares of real numbers at every point of $\calZ_d$, so each $q_{\alpha}^{[d]}$ vanishes there.
    Proposition~\ref{prop:vanishing-structure} therefore gives polynomials $h_{\alpha,i,j}\in\bR[\sX,\sY]$ such that
    \begin{align}
        q_{\alpha}^{[d]} &= \sum_{1\le i, j \le n} h_{\alpha,i,j} \cdot (\sX\sY^\top)_{i,j}, \quad \text{ for } 1\le \alpha \le 2m,\label{eq:q-structure}
    \end{align}
    and $\deg(h_{\alpha,i,j}) \le d-2$.
    Since $q_{\alpha}$ has degree at most $d$, we have $q_{\alpha} = q_{\alpha}^{[d]} + \hat{q}_{\alpha}$, where
    \begin{align*}
        \hat{q}_{\alpha} = \sum_{0 \le i < d} q_{\alpha}^{[i]}
    \end{align*}
    has degree at most $d-1$. Substituting $q_{\alpha} = q_{\alpha}^{[d]} + \hat{q}_{\alpha}$ and \eqref{eq:q-structure} into \eqref{eq:min-deg-SoS-WRank}, we get
    \begin{align*}
        -1 &= \sum_{\alpha = 1}^{2m} \left(\hat{q}_{\alpha} + \sum_{1\le i, j \le n} h_{\alpha,i,j} \cdot (\sX\sY^\top)_{i,j}\right)^2 + \sum_{1\le i,j\le n} h_{i,j}\cdot \left((\sX\sY^\top)_{i,j} - \delta_{i,j}\right)\\
        &= \sum_{\alpha = 1}^{2m} \left(\hat{q}_{\alpha} + \sum_{i=1}^n h_{\alpha,i,i}\right)^2 + \sum_{\alpha = 1}^{2m} \left(\left(\hat{q}_{\alpha} + \sum_{1\le i, j \le n} h_{\alpha,i,j} \cdot (\sX\sY^\top)_{i,j}\right)^2-\left(\hat{q}_{\alpha} + \sum_{i=1}^n h_{\alpha,i,i}\right)^2\right) \\
        &\qquad + \sum_{1\le i,j\le n} h_{i,j}\cdot \left((\sX\sY^\top)_{i,j} - \delta_{i,j}\right)\\
        &= \sum_{\alpha = 1}^{2m} \left(\hat{q}_{\alpha} + \sum_{i=1}^n h_{\alpha,i,i}\right)^2 + \sum_{1\le i,j\le n} h_{i,j}\cdot \left((\sX\sY^\top)_{i,j} - \delta_{i,j}\right) \\
        &\qquad + \sum_{\alpha = 1}^{2m} \left(2\hat{q}_{\alpha} + \sum_{1\le i, j \le n} h_{\alpha,i,j} \cdot \left((\sX\sY^\top)_{i,j} + \delta_{i,j}\right)\right)\left(\sum_{1\le i, j \le n} h_{\alpha,i,j} \cdot \left((\sX\sY^\top)_{i,j} - \delta_{i,j}\right)\right)\\
        &= \sum_{\alpha = 1}^{2m} \left(\hat{q}_{\alpha} + \sum_{i=1}^n h_{\alpha,i,i}\right)^2 + \sum_{1\le i,j\le n} \hat{h}_{i,j}\cdot \left((\sX\sY^\top)_{i,j} - \delta_{i,j}\right), \numberthis \label{eq:smaller-deg-SoS-WRank}
    \end{align*}
    where for $1\le i,j\le n$ we denote
    \begin{align*}
        \hat{h}_{i,j} &\colonequals h_{i,j} + \sum_{\alpha=1}^{2m} h_{\alpha,i,j} \left(2\hat{q}_{\alpha} + \sum_{1\le k,\ell \le n} h_{\alpha,k,\ell} \cdot \left((\sX\sY^\top)_{k,\ell} + \delta_{k,\ell}\right)\right).
    \end{align*}

    The polynomials inside the new squares satisfy $\deg\left(\hat{q}_{\alpha} + \sum_{i=1}^n h_{\alpha,i,i}\right) \le \max\{d-1, d-2\} \le d-1$.
    Also,
    \begin{align*}
        \deg\left(2\hat{q}_{\alpha} + \sum_{1\le k,\ell \le n} h_{\alpha,k,\ell} \cdot \left((\sX\sY^\top)_{k,\ell} + \delta_{k,\ell}\right)\right) &\le \max\left\{\deg(q_{\alpha}), \deg\left(\hat{q}_{\alpha} + \sum_{i=1}^n h_{\alpha,i,i}\right)\right\} \le d,
    \end{align*}
    Since $\deg(h_{i,j})\le D-2$, $\deg(h_{\alpha,i,j})\le d-2$, and $2d\le D$, we obtain
    \begin{align*}
        \deg(\hat{h}_{i,j}) \le \max\{D-2, (d-2)+d\}  = D-2.
    \end{align*}

    Thus \eqref{eq:smaller-deg-SoS-WRank} is a degree-$D$ SoS proof whose sum-of-squares term
    \[\sum_{\alpha = 1}^{2m} \left(\hat{q}_{\alpha} + \sum_{i=1}^n h_{\alpha,i,i}\right)^2\]
    has degree at most $2d-2$, contradicting the minimality of $d$.
\end{proof}

\section{SoS Complexity of $\TFrame$, $\Inv_n$, and $\Orthog_n$}

\subsection{Tight Frame and Orthogonality Principles: Proofs of Theorems~\ref{thm:TFrame-thm} and \ref{thm:Orth-thm}}

We show that unlike $\WRank$, $\TFrame_r^n$ has a degree-$4$ SoS proof when $n > r$. We also prove that $\Orthog_n$ has a degree-$4$ SoS proof. Our proofs rely on the following identity.

\begin{lemma}\label{lem:quartic-identity}
    Let $n$ and $r$ be positive integers. Let $\sX$ be an $n \times r$ matrix of real indeterminates. Then,
    \begin{align*}
        \sum_{i,j=1}^r \left((\sX^\top \sX)_{i,j} - \delta_{i,j}\right)^2 - r = \sum_{i,j=1}^n \left((\sX \sX^\top)_{i,j} - \delta_{i,j}\right)^2 - n.
    \end{align*}
\end{lemma}

\begin{proof}
    Observe the following polynomial identities:
    \begin{align*}
        \tr((\sX^\top \sX)^2) &= \tr((\sX \sX^\top)^2),\\
        \tr(\sX^\top \sX) &= \tr(\sX \sX^\top).
    \end{align*}
    Therefore, it holds that
    \begin{align*}
        &\sum_{i,j=1}^r \left((\sX^\top \sX)_{i,j} - \delta_{i,j}\right)^2 - r = \sum_{i,j=1}^r (\sX^\top \sX)_{i,j}^2 - 2\sum_{i=1}^r (\sX^\top \sX)_{i,i} + r - r\\
        &= \tr((\sX \sX^\top)^2) - 2\tr(\sX \sX^\top) = \tr((\sX^\top \sX)^2) - 2\tr(\sX^\top \sX)\\
        &= \sum_{i,j=1}^n (\sX \sX^\top)_{i,j}^2 - 2\sum_{i=1}^n (\sX \sX^\top)_{i,i} + n - n = \sum_{i,j=1}^n \left((\sX \sX^\top)_{i,j} - \delta_{i,j}\right)^2 - n.
    \end{align*}
\end{proof}

Now we prove that $\TFrame$ and $\Orthog$ has a degree-$4$ SoS proof.
\begin{proof}[Proof of Theorem~\ref{thm:TFrame-thm}]
    By Lemma~\ref{lem:quartic-identity}, for $n \times r$ matrix $\sX$, we have
    \begin{align*}
        -1 = \frac{1}{n-r}\left(\sum_{i,j=1}^r \left((\sX^\top \sX)_{i,j} - \delta_{i,j}\right)^2 - \sum_{i,j=1}^n \left((\sX \sX^\top)_{i,j} - \delta_{i,j}\right)^2\right),
    \end{align*}
    which is a degree-$4$ SoS refutation of $\sX\sX^\top = \II_n$ when $n > r$.
\end{proof}

\begin{proof}[Proof of Theorem~\ref{thm:Orth-thm}]
    By Lemma~\ref{lem:quartic-identity}, for $n \times n$ matrix $\sX$, we have
    \begin{align*}
        \sum_{i,j=1}^n \left((\sX^\top \sX)_{i,j} - \delta_{i,j}\right)^2 = \sum_{i,j=1}^n \left((\sX \sX^\top)_{i,j} - \delta_{i,j}\right)^2,
    \end{align*}
    which is an SoS proof of degree $4$ of the implication $\sX\sX^\top = \II_n \Longrightarrow \sX^\top \sX = \II_n$.
\end{proof}

\subsection{Inversion Principle: Proof of Theorem~\ref{thm:Inv-thm}}
In this section, we prove Theorem~\ref{thm:Inv-thm}, showing that $\Inv_n$ needs an SoS degree of $2n+2$.

\begin{proof}[Proof of Theorem~\ref{thm:Inv-thm}]
    First let us consider $n = 1$. In this case $\sX$ and $\sY$ are simply real indeterminates, and both the premise and the conclusion of $\Inv_1$ are $\sX\sY = 1$. It is easy to check that degree $4$ is necessary and sufficient for an SoS proof of $\Inv_1$.

    Now we fix $n \ge 2$. Since there does not exist an SoS proof of $\WRank_n^{n+1}(\II_{n+1})$ of degree at most $2n+1$ by Theorem~\ref{thm:WRank-thm}, there does not exist an SoS proof of $\Inv_n$ of degree at most $2n+1$ by Proposition~\ref{prop:inversion-rank-reduction}.

    Next we construct an SoS proof of $\Inv_n$ of degree $2n+2$. Let $\uu,\vv \in \bR^n$ be constant vectors. We form a matrix
    \begin{align*}
        \sM \colonequals \begin{pmatrix}
            \sX\sY^\top & \sX\uu\\
            \vv^\top\sY^\top & \vv^\top\uu
        \end{pmatrix}.
    \end{align*}
    Note that $\sM$ can be factorized as a product of $(n+1) \times n$ matrix and $n \times (n+1)$ matrix
    \begin{align*}
        \sM = \begin{pmatrix}
            \sX\\
            \vv^\top
        \end{pmatrix} \begin{pmatrix}
            \sY^\top & \uu
        \end{pmatrix},
    \end{align*}
    and thus $\det(\sM) = 0$. We also consider the following $(n+1) \times (n+1)$ matrix $\sB$ which replaces the top-left block of $\sM$ by $\II_n$:
    \begin{align*}
        \sB \colonequals \begin{pmatrix}
            \II_n & \sX\uu\\
            \vv^\top\sY^\top & \vv^\top\uu
        \end{pmatrix}.
    \end{align*}

    For $0 \le j \le n$, let $\sM^{(j)}$ be the $(n+1) \times (n+1)$ matrix with its first $j$ columns from $\sM$ and its last $n+1-j$ columns from $\sB$. Then, $\sM^{(n)} = \sM$ and $\sM^{(0)} = \sB$. Since the matrices $\sM^{(j)}$ and $\sM^{(j-1)}$ differ only in column $j$, expanding their determinants along this column gives
    \[
        \det(\sM^{(j)})-\det(\sM^{(j-1)})
        =\sum_{i=1}^n H_{i,j}(\sM_{i,j}-\sB_{i,j}),
    \]
    where $H_{i,j}$ is the $(i,j)$ cofactor of $\sM^{(j-1)}$.
    Each cofactor is, up to sign, a determinant of size $n$ whose entries are constants, linear polynomials, or quadratic polynomials in $\sX,\sY$. We note that every term appearing in the $(i,j)$ cofactor of $\sM^{(j-1)}$ is either a product of $n-1$ entries in the top-left block of $\sM^{(j-1)}$ and the constant entry $\vv^\top \uu$, or a product of $n-2$ entries in the top-left block of $\sM^{(j-1)}$, one degree-$1$ entry in $\vv^\top \sY^\top$, and one degree-$1$ entry in $\sX\uu$. Since every entry in the top-left block of $\sM^{(j-1)}$ has degree at most $2$, we conclude that $H_{i,j}$ has degree at most $2n-2$. 

    Summing over $j$ and using $\det(\sM) = 0$ gives
    \begin{align*}
        -\det(\sB) = \sum_{i,j=1}^n H_{i,j}(\sM_{i,j} - \sB_{i,j}).
    \end{align*}
    Note that $\sM_{i,j} - \sB_{i,j} = (\sX\sY^\top)_{i,j} - \delta_{i,j}$ for $1\le i,j\le n$, and $\det(\sB)$ can be expanded as
    \begin{align*}
        \det(\sB) &= \det\begin{pmatrix}
            \II_n & \sX\uu\\
            \vv^\top\sY^\top & \vv^\top\uu
        \end{pmatrix} = \det\begin{pmatrix}
            \II_n & \sX\uu\\
            \zzero & \vv^\top\uu - \vv^\top \sY^\top \sX \uu
        \end{pmatrix} = \vv^\top\uu - \vv^\top \sY^\top \sX \uu.
    \end{align*}
    Thus, for $\uu, \vv \in \bR^n$, we have
    \begin{align*}
        \vv^\top \left(\sY^\top \sX - \II_n\right) \uu = \sum_{i,j=1}^n H_{i,j}^{(\uu,\vv)}\left((\sX\sY^\top)_{i,j} - \delta_{i,j}\right).
    \end{align*}
    Now for $1\le a,b\le n$, we set $\vv = \ee_a, \uu = \ee_b$ and get
    \begin{align}
        (\sY^\top \sX)_{a,b} - \delta_{a,b} = \sum_{i,j=1}^n H_{i,j}^{(\ee_b,\ee_a)}\left((\sX\sY^\top)_{i,j} - \delta_{i,j}\right). \label{eq:u-v-identity}
    \end{align}
    Taking the squares and summing them up, we get
    \begin{align*}
        \sum_{a,b=1}^n \left((\sY^\top \sX)_{a,b} - \delta_{a,b}\right)^2 =  \sum_{i,j=1}^n \left((\sX\sY^\top)_{i,j} - \delta_{i,j}\right) \sum_{a,b=1}^n H_{i,j}^{(\ee_b,\ee_a)}\left((\sY^\top \sX)_{a,b} - \delta_{a,b}\right).
    \end{align*}
    Note that the summands on the right hand side have degree at most $2 + 2n-2 + 2 = 2n+2$, and this gives an SoS proof of $\Inv_n$ of degree $2n+2$.
\end{proof}

\section{SoS Complexity of $\NormWRank$}

In this section, we prove Theorem~\ref{thm:norm-constraints-upper-bound} and Theorem~\ref{thm:norm-constraints-lower-bound}, showing how the SoS complexity of Norm-Constrained Weak Rank Principle changes as the norm budget $B$ changes.

\subsection{Upper Bound: Proof of Theorem~\ref{thm:norm-constraints-upper-bound}}

The following identity is a simple consequence of the cyclicity of trace.

\begin{lemma}\label{lem:polynomial-trace-transpose-identity}
    Let $n > r$ be positive integers, and $\sZ$ be an $n \times r$ matrix of real indeterminates. Then, for every univariate polynomial $f(z)$,
    \begin{align*}
        \tr(f(\sZ^\top\sZ)) = \tr(f(\sZ\sZ^\top)) - (n-r)f(0).
    \end{align*}
\end{lemma}

\begin{proof}
    By cyclicity of trace, for every integer $i \ge 1$, 
    \begin{align*}
        \tr((\sZ^\top\sZ)^i) = \tr((\sZ\sZ^\top)^i).
    \end{align*}
    For $i = 0$, we have
    \begin{align*}
        \tr((\sZ^\top\sZ)^0) = \tr(\II_r) = \tr(\II_n) - (n-r) = \tr((\sZ\sZ^\top)^0) - (n-r).
    \end{align*}
    Applying these identities to the polynomial $f(z) = \sum_{i=0}^d a_i z^i$, we get
    \begin{align*}
        \tr(f(\sZ^\top\sZ)) &= \sum_{i=0}^d a_i \tr((\sZ^\top\sZ)^i)\\
        &= - (n-r) a_0 + \sum_{i=0}^d a_i \tr((\sZ\sZ^\top)^i)\\
        &= \tr(f(\sZ\sZ^\top)) - (n-r)f(0),
    \end{align*}
    since $a_0 = f(0)$.
\end{proof}

Now we prove the SoS upper bound for the Norm-Constrained Weak Rank Principle stated in Theorem~\ref{thm:norm-constraints-upper-bound}. The high level idea of the proof is as follows. If we set $\sU = \frac{\sX+\sY}{2}$ and $\sV = \frac{\sX-\sY}{2}$, then modulo the constraint equations we have $\sU\sU^\top - \II_n \equiv \sV\sV^\top$ and $\|\sV\|_F^2 + \frac{\st^2}{4} \equiv \frac{B-2n}{4}$. The first identity contains the rank obstruction as $\sU\sU^\top - \II_n$ should have $n-r$ eigenvalues of $-1$ whereas $\sV\sV^\top$ is positive semidefinite, and the second identity says that the Frobenius norm of the difference $\sX - \sY$ is controlled by $B-2n$. Using the trace identity in Lemma~\ref{lem:polynomial-trace-transpose-identity}, we could show that for a univariate polynomial $p(z)$ with $p(0) = 0$, we have
\begin{align*}
    \|p(\sU^\top\sU - \II_r)\|_F^2 \equiv \|p(\sV^\top \sV)\|_F^2 - (n-r)p(-1)^2
\end{align*}
modulo the constraints. If hypothetically the eigenvalues of $\sV^\top \sV$ are $\lambda_j \ge 0$, then $\|\sV\|_F^2 + \frac{\st^2}{4} \equiv \frac{B-2n}{4}$ should force that $\sum_{j} \lambda_j \le \frac{B-2n}{4}$. To construct an SoS refutation, our goal is then to choose a univariate polynomial $p(z)$ that satisfies $p(0) = 0$, $p(z)$ is moderate for $0 \le z \le \frac{B-2n}{4}$, and $p(-1)^2$ is large, so that the right hand side is negative whereas the left hand side is nonnegative. Our construction of $p(z)$ makes use of Chebyshev polynomials, which have the extremal property of being bounded in $[-1,1]$ and growing rapidly outside it.

\begin{proof}[Proof of Theorem~\ref{thm:norm-constraints-upper-bound}]
    Let $\sX,\sY$ be $n \times r$ matrices of real indeterminates. Denote $\sU \colonequals \frac{\sX+ \sY}{2}$ and $\sV \colonequals \frac{\sX - \sY}{2}$. Then, we have
    \begin{align*}
        \sU\sU^\top - \sV\sV^\top - \II_n  &= \frac{(\sX\sY^\top - \II_n) + (\sX\sY^\top - \II_n)^\top}{2}, \numberthis \label{eq:U-V-cong}\\
        \frac{B - 2n}{4} - \|\sV\|_F^2&= \frac{B}{4} - \frac{\|\sX\|_F^2 + \|\sY\|_F^2 - 2\tr(\sX\sY^\top)}{4} - \frac{n}{2}\\
        &= \frac{\st^2}{4} - \frac{\|\sX\|_F^2 + \|\sY\|_F^2 + \st^2 - B}{4} + \frac{1}{2}\sum_{i=1}^n \left((\sX\sY^\top)_{i,i} - 1\right). \numberthis\label{eq:B-V-cong}
    \end{align*}
    Thus, modulo the polynomial constraints $\sX\sY^\top = \II_n$ and $\|\sX\|_F^2 + \|\sY\|_F^2 + \st^2 = B$, we have $\sU\sU^\top - \II_n \equiv \sV\sV^\top$ and $\frac{B-2n}{4} - \|\sV\|_F^2 \equiv \frac{\st^2}{4}$.

    Now consider an arbitrary univariate polynomial $f(z) = \sum_{k=0}^m a_k z^k$ of degree $m$. Applying Lemma~\ref{lem:polynomial-trace-transpose-identity} to $\sZ = \sU$ and $f(z-1)$, we have
    \begin{align*}
        &\tr f(\sU^\top \sU - \II_r)\\
        &= \tr f(\sU\sU^\top - \II_n) - (n-r)f(-1)
        \intertext{If we denote $\sM \colonequals \sV\sV^\top + \frac{(\sX\sY^\top - \II_n) + (\sX\sY^\top - \II_n)^\top}{2}$, then $\sU\sU^\top - \II_n = \sM$ by \eqref{eq:U-V-cong} and we get}
        &= \tr f(\sV\sV^\top) - (n-r)f(-1) + \tr\left(f\left(\sM\right) -  f(\sV\sV^\top)\right)\\
        &= \tr f(\sV\sV^\top) - (n-r)f(-1) + \sum_{k=1}^m a_k \tr\left(\sM^k - (\sV\sV^\top)^k\right)
        \intertext{Since $\sM^k - (\sV\sV^\top)^k = \sum_{\ell = 0}^{k-1}\sM^{k-1-\ell} \frac{(\sX\sY^\top - \II_n) + (\sX\sY^\top - \II_n)^\top}{2}(\sV\sV^\top)^{\ell}$, every entry of $\sM^k - (\sV\sV^\top)^k$ is a polynomial combination of the constraint equations $\sX\sY^\top - \II_n$, and there exist polynomials $h_{k,i,j} \in \bR[\sX,\sY]$ of degree at most $2k-2$ such that $\tr\left(\sM^k - (\sV\sV^\top)^k\right) = \sum_{1\le i,j\le n} h_{k,i,j}\cdot \left((\sX\sY^{\top})_{i,j} - \delta_{i,j}\right)$. Setting $h_{i,j}^{(f)} \colonequals \sum_{k=1}^m a_k h_{k,i,j}$ for $1\le i,j\le n$ which has degree at most $2m-2$, we get}
        &= \tr f(\sV\sV^\top) - (n-r)f(-1) + \sum_{1\le i,j\le n} h_{i,j}^{(f)}\cdot \left((\sX\sY^{\top})_{i,j} - \delta_{i,j}\right)
        \intertext{Now we apply Lemma~\ref{lem:polynomial-trace-transpose-identity} to $\sZ = \sV$ and $f(z)$ and get}
        &= \tr f(\sV^\top \sV) + (n-r)(f(0) - f(-1)) + \sum_{1\le i,j\le n} h_{i,j}^{(f)}\cdot \left((\sX\sY^{\top})_{i,j} - \delta_{i,j}\right).
    \end{align*}

    If we specialize the above identity to $f(z) = p(z)^2$ for some univariate polynomial $p(z)$, we get
    \begin{align*}
        \|p(\sU^\top\sU - \II_r)\|_F^2 = \|p(\sV^\top \sV)\|_F^2 + (n-r)(p(0)^2 - p(-1)^2) + \sum_{1\le i,j\le n} h_{i,j}^{(p^2)}\cdot \left((\sX\sY^{\top})_{i,j} - \delta_{i,j}\right). \numberthis\label{eq:p-Frobenius-identity}
    \end{align*}

    Now we construct a polynomial $p(z)$ that leads to an SoS refutation. Recall that the Chebyshev polynomials of the first kind are given by $T_j(\cos \theta) \colonequals \cos(j\theta)$ where $j \in \bN$ and the Chebyshev polynomials of the second kind of are given by $U_j(\cos \theta) \colonequals \frac{\sin((j+1)\theta)}{\sin \theta}$ where $j \in \bN$. The two kinds of Chebyshev polynomials are related by Pell equation:
    \begin{align*}
        1 - T_j(z)^2 &= (1 - z^2) U_{j-1}(z)^2, \text{ for } j \ge 1.
    \end{align*}

    We set $p(z) \colonequals \sqrt{z} T_{2D-1}\left(2\sqrt{\frac{z}{B-2n}}\right)$ as a univariate polynomial of degree $D$ and $q(z) \colonequals U_{2D-2}\left(2\sqrt{\frac{z}{B-2n}}\right)$ as a univariate polynomial of degree $D-1$, which are well-defined since $T_{2D-1}(z)$ is an odd polynomial and $U_{2D-2}(z)$ is an even polynomial. By Pell equation, we have
    \begin{align*}
        z - p(z)^2 &= z\left(1 - T_{2D-1}\left(2\sqrt{\frac{z}{B-2n}}\right)^2\right)\\
        &= \frac{z(B-2n-4z)}{B-2n} U_{2D-2}\left(2\sqrt{\frac{z}{B-2n}}\right)^2 = \frac{z(B-2n-4z)}{B-2n} q(z)^2. \numberthis \label{eq:pell-eq}
    \end{align*}

    By Lagrange's identity,
    \begin{align*}
        &\|\sV\|_F^2 \|\sV q(\sV^\top \sV)\|_F^2 - \|\sV^\top \sV q(\sV^\top \sV)\|_F^2\\
        &= \left(\sum_{i=1}^n \sum_{a = 1}^r \sV_{i,a}^2\right) \left(\sum_{i=1}^n \sum_{b = 1}^r (\sV q(\sV^\top \sV))_{i,b}^2\right) - \sum_{a = 1}^r \sum_{b=1}^r\left(\sum_{i=1}^n \sV_{i,b}(\sV q(\sV^\top \sV))_{i,a}\right)^2\\
        &= \sum_{1\le a,b\le r} \sum_{1\le i< j\le n} \left(\sV_{i,a}(\sV q(\sV^\top \sV))_{j,b} - \sV_{j,a}(\sV q(\sV^\top \sV))_{i,b}\right)^2.
    \end{align*}
    Therefore, substituting $z = \sV^\top \sV$ in \eqref{eq:pell-eq} and taking traces gives
    \begin{align*}
        &\|\sV\|_F^2 - \|p(\sV^\top \sV)\|_F^2\\ &= \|\sV q(\sV^\top \sV)\|_F^2 - \frac{4}{B-2n}\|\sV^\top \sV q(\sV^\top \sV)\|_F^2\\
        &= \left(1 - \frac{4}{B-2n}\|\sV\|_F^2\right) \|\sV q(\sV^\top \sV)\|_F^2 + \frac{4}{B-2n}\left(\|\sV\|_F^2 \|\sV q(\sV^\top \sV)\|_F^2 - \|\sV^\top \sV q(\sV^\top \sV)\|_F^2\right)\\
        &= \left(1 - \frac{4}{B-2n}\|\sV\|_F^2\right) \|\sV q(\sV^\top \sV)\|_F^2\\
        &\hspace{0.4in} + \frac{4}{B-2n} \sum_{1\le a,b\le r} \sum_{1\le i< j\le n} \left(\sV_{i,a}(\sV q(\sV^\top \sV))_{j,b} - \sV_{j,a}(\sV q(\sV^\top \sV))_{i,b}\right)^2.
    \end{align*}

    Substituting $\|\sV\|_F^2$ in the identity above using \eqref{eq:B-V-cong}, we get
    \begin{align*}
        &\frac{B-2n}{4} - \|p(\sV^\top \sV)\|_F^2\\
        &= \frac{\st^2}{4} - \frac{\|\sX\|_F^2 + \|\sY\|_F^2 + \st^2 - B}{4} + \frac{1}{2}\sum_{i=1}^n \left((\sX\sY^\top)_{i,i} - 1\right)\\
        &\hspace{0.4in} + \frac{4}{B-2n}\left(\frac{\st^2}{4} - \frac{\|\sX\|_F^2 + \|\sY\|_F^2 + \st^2 - B}{4} + \frac{1}{2}\sum_{i=1}^n \left((\sX\sY^\top)_{i,i} - 1\right)\right) \|\sV q(\sV^\top \sV)\|_F^2\\
        &\hspace{0.4in}+ \frac{4}{B-2n} \sum_{1\le a,b\le r} \sum_{1\le i< j\le n} \left(\sV_{i,a}(\sV q(\sV^\top \sV))_{j,b} - \sV_{j,a}(\sV q(\sV^\top \sV))_{i,b}\right)^2\\
        &= \frac{\st^2}{4} + \frac{\st^2}{B-2n} \|\sV q(\sV^\top \sV)\|_F^2 +\frac{4}{B-2n} \sum_{1\le a,b\le r} \sum_{1\le i< j\le n} \left(\sV_{i,a}(\sV q(\sV^\top \sV))_{j,b} - \sV_{j,a}(\sV q(\sV^\top \sV))_{i,b}\right)^2\\
        &\hspace{0.4in} - \left(\frac{1}{4} + \frac{1}{B-2n}\|\sV q(\sV^\top \sV)\|_F^2\right)(\|\sX\|_F^2 + \|\sY\|_F^2 + \st^2 - B)\\
        &\hspace{0.4in} + \frac{1}{2}\left(1 + \|\sV q(\sV^\top \sV)\|_F^2\right)\sum_{i=1}^n \left((\sX\sY^\top)_{i,i} - 1\right).
    \end{align*}
    Rearranging the equation and substituting $\|p(\sV^\top \sV)\|_F^2$ using \eqref{eq:p-Frobenius-identity}, we get
    \begin{align*}
        &\frac{B-2n}{4} - (n-r)\left(p(-1)^2 - p(0)^2\right)\\
        &= \|p(\sU^\top \sU - \II_r)\|_F^2 + \frac{\st^2}{4} + \frac{\st^2}{B-2n} \|\sV q(\sV^\top \sV)\|_F^2 \\
        &\hspace{0.4in}+\frac{4}{B-2n} \sum_{1\le a,b\le r} \sum_{1\le i< j\le n} \left(\sV_{i,a}(\sV q(\sV^\top \sV))_{j,b} - \sV_{j,a}(\sV q(\sV^\top \sV))_{i,b}\right)^2\\
        &\hspace{0.4in} - \left(\frac{1}{4} + \frac{1}{B-2n}\|\sV q(\sV^\top \sV)\|_F^2\right)(\|\sX\|_F^2 + \|\sY\|_F^2 + \st^2 - B)\\
        &\hspace{0.4in} + \sum_{i=1}^n \left(\frac{1}{2} + \frac{1}{2}\|\sV q(\sV^\top \sV)\|_F^2 + h_{i,i}^{(p^2)}\right)\left((\sX\sY^\top)_{i,i} - 1\right) + \sum_{\substack{1\le i,j\le n:\\ i\ne j} } h_{i,j}^{(p^2)}\cdot (\sX\sY^\top)_{i,j}, \numberthis \label{eq:SoS-refutation-NormWRank}
    \end{align*}
    which expresses the left hand side constant $\frac{B-2n}{4} - (n-r)(p(-1)^2 - p(0)^2)$ as a sum-of-squares polynomial modulo the constraints $\|\sX\|_F^2 + \|\sY\|_F^2 + \st^2 = B$ and $\sX\sY^\top = \II_n$. Recall that $p(z) = \sqrt{z}\cdot  T_{2D-1}\left(2\sqrt{\frac{z}{B-2n}}\right)$. Thus, $p(0) = 0$. On the other hand, $T_{2D-1}(\cos \theta) = \cos((2D-1)\theta)$ extends to complex $\theta$ by analyticity, and we have $T_{2D-1}(\cosh(\theta)) = \cosh((2D-1)\theta)$ for any $\theta \in \bC$ using that $\cos(\imag\theta) = \cosh(\theta)$ where $\imag$ is the imaginary unit. Set $u \colonequals \arcsinh\left(\frac{2}{\sqrt{B-2n}}\right)$. We may evaluate
    \begin{align*}
        p(-1)^2 &= - T_{2D-1}\left(2\sqrt{-\frac{1}{B-2n}}\right)^2\\
        &= -T_{2D-1}(\imag \sinh(u))^2\\
        &= -T_{2D-1}\left(\cosh\left(u + \imag\frac{\pi}{2}\right)\right)^2\\
        &= -\cosh^2\left((2D-1)\left(u + \imag \frac{\pi}{2}\right)\right)\\
        &= \sinh^2((2D-1)u), 
    \end{align*}
    where we used that $\imag\sinh(u) = \cosh(u + \imag \pi/2)$ and $\imag\cosh(u) = \sinh(u + \imag \pi/2)$.

    By the assumption that
    \[(n-r)\sinh^2\left((2D-1)\arcsinh\left(\frac{2}{\sqrt{B-2n}}\right)\right) > \frac{B-2n}{4},\]
    the left hand side constant of \eqref{eq:SoS-refutation-NormWRank}
    \begin{align*}
        &\frac{B-2n}{4} - (n-r)\left(p(-1)^2 - p(0)^2\right)\\
        &= \frac{B-2n}{4} - (n-r)\sinh^2\left((2D-1)\arcsinh\left(\frac{2}{\sqrt{B-2n}}\right)\right)
    \end{align*}
    is negative. Thus, after the left hand side is normalized to $-1$, \eqref{eq:SoS-refutation-NormWRank} is an SoS proof of the infeasibility of the polynomial system of $\NormWRank_r^{n,B}(\II_n)$. Note that every term on the right hand side of \eqref{eq:SoS-refutation-NormWRank} has degree at most $4D$, since $p(z)$ has degree $D$ and $q(z)$ has degree $D-1$.
\end{proof}

Next, we derive Corollary~\ref{cor:norm-constraints-upper-bound}, which says that an SoS proof of $\NormWRank_r^{n,B}(\II_r)$ exists at degree $O(\sqrt{n}\log(n))$ whenever the norm budget satisfies $B = O(n)$.

\begin{proof}[Proof of Corollary~\ref{cor:norm-constraints-upper-bound}]
    Fix $C > 2$ and $2n < B < Cn$. By Theorem~\ref{thm:norm-constraints-upper-bound}, for any integer $D \ge 1$, if
    \begin{align*}
        (n-r)\sinh^2\left((2D-1)\arcsinh\left(\frac{2}{\sqrt{B-2n}}\right)\right) > \frac{B-2n}{4},
    \end{align*}
    then there exists a degree-$4D$ SoS proof of $\NormWRank_r^{n,B}(\II_n)$. Equivalently, we need
    \begin{align*}
        2D-1 \ge \frac{\arcsinh\left(\frac{\sqrt{B-2n}}{2(n-r)}\right)}{\arcsinh\left(\frac{2}{\sqrt{B-2n}}\right)}. \numberthis\label{ineq:D-condition}
    \end{align*}
    
    Using that $\arcsinh(z) = \log(z + \sqrt{z^2 + 1})$ for $z \in \bR$ and $2n < B < Cn$, we have
    \begin{align*}
        \arcsinh\left(\frac{\sqrt{B-2n}}{2(n-r)}\right) &= \log\left(\frac{\sqrt{B-2n}}{2(n-r)} + \sqrt{1 + \frac{B-2n}{4(n-r)^2}}\right) \le O_C(\log(n)).\\
        \arcsinh\left(\frac{2}{\sqrt{B-2n}}\right) &= \log\left(\frac{2}{\sqrt{B-2n}} + \sqrt{1 + \frac{4}{B-2n}}\right) \ge \Omega_C\left(\frac{1}{\sqrt{n}}\right).
    \end{align*}
    Therefore, to satisfy \eqref{ineq:D-condition}, it suffices to take $D = O_C\left(\sqrt{n}\log(n)\right)$.
\end{proof}

\subsection{Lower Bound for Large $B$: Proof of Theorem~\ref{thm:norm-constraints-lower-bound}}

First, we establish the following auxiliary results.

\begin{proposition}\label{prop:SoS-structure}
    Let $r < n$ be positive integers. Let $\sX, \sY$ be $n \times r$ matrices of real indeterminates, and $\sum_{\alpha=1}^m p_\alpha^2 \in \bR[\sX, \sY]$ be a sum-of-squares polynomial of degree at most $2r$. If there exist polynomials $h_{i,j} \in \bR[\sX, \sY]$ for $1 \le i,j\le n$ of degrees at most $2r-2$ such that
    \begin{align*}
        \sum_{\alpha=1}^m p_\alpha^2 = \sum_{1\le i, j\le n} h_{i,j} \cdot \left((\sX\sY^\top)_{i,j} - \delta_{i,j}\right),
    \end{align*}
    then for every $1\le \alpha \le m$, there exist polynomial $h_{\alpha,i,j} \in  \bR[\sX, \sY]$ for $1 \le i,j\le n$ of degrees at most $\deg(p_{\alpha})-2$ such that
        \begin{align*}
            p_{\alpha} = \sum_{1\le i, j\le n} h_{\alpha,i,j} \cdot \left((\sX\sY^\top)_{i,j} - \delta_{i,j}\right).
        \end{align*}
\end{proposition}

\begin{proof}
    We prove the claim by induction on the maximum degree $d \colonequals \max_{1\le \alpha \le m} \deg(p_{\alpha})$.

    If $d \le 0$, the every $p_{\alpha}$ is a constant polynomial. By Proposition~\ref{prop:vanishing-top-degree}, $p_{\alpha}$ vanishes on $\calZ_0 \colonequals \{(\zzero, \zzero)\}$. Thus, we have $p_{\alpha} = 0$ for every $1\le \alpha \le m$.

    Fix $1\le d \le r$ and suppose that we have proved the claim for every $d' < d$. By Proposition~\ref{prop:vanishing-top-degree}, $(\sum_{\alpha = 1}^m p_{\alpha}^2)^{[2d]}$ vanishes on $\calZ_d$. Therefore, for $1\le \alpha \le m$ such that $\deg(p_{\alpha}) = d$, $p_{\alpha}^{[d]}$ vanishes on $\calZ_d$, and by Proposition~\ref{prop:vanishing-structure} there exist polynomial $h_{\alpha,i,j} \in \bR[\sX, \sY]$ for $1\le i,j\le n$ of degrees at most $d-2$ such that
    \begin{align*}
        p_{\alpha}^{[d]} = \sum_{1\le i,j\le n} h_{\alpha,i,j} \cdot (\sX\sY^\top)_{i,j}. \numberthis \label{eq:p-alpha-ideal}
    \end{align*}
    
    For $1\le \alpha\le m$ such that $\deg(p_{\alpha}) \le d-1$, we set $\Tilde{p}_{\alpha} = p_{\alpha}$. For $1\le \alpha\le m$ such that $\deg(p_{\alpha}) = d$, we define $\Tilde{p}_{\alpha} = p_{\alpha} - p_{\alpha}^{[d]} + \sum_{i=1}^n h_{\alpha,i,i}$, and note that
    \begin{align*}
        p_{\alpha}^2 - \Tilde{p}_{\alpha}^2 &= (p_{\alpha} + \Tilde{p}_{\alpha})(p_{\alpha}^{[d]} - \sum_{i=1}^n h_{\alpha,i,i}) = (p_{\alpha} + \Tilde{p}_{\alpha})\sum_{1\le i,j\le n} h_{\alpha,i,j} \cdot \left((\sX\sY^\top)_{i,j} - \delta_{i,j}\right),
    \end{align*}
    where $(p_{\alpha} + \Tilde{p}_{\alpha})h_{\alpha,i,j}$ has degree at most $2d-2$. Therefore, we may apply the inductive hypothesis to the sum-of-squares polynomial
    \begin{align*}
        \sum_{\alpha = 1}^m \Tilde{p}_{\alpha}^2 &= \sum_{\alpha=1}^m p_{\alpha}^2 - \sum_{\alpha=1}^m \left(p_{\alpha}^2 - \Tilde{p}_{\alpha}^2\right)\\
        &= \sum_{1\le i,j\le n} \left(h_{i,j} - \sum_{\alpha=1}^m (p_{\alpha} + \Tilde{p}_{\alpha})h_{\alpha,i,j}\right) \cdot \left((\sX\sY^\top)_{i,j} - \delta_{i,j}\right)
    \end{align*}
    where $\deg(\Tilde{p}_{\alpha}) \le d-1$, to show that there exists $\Tilde{h}_{\alpha, i, j} \in \bR[\sX,\sY]$ for $1\le i,j \le n$ of degrees at most $\deg(\Tilde{p}_{\alpha}) - 2$ such that $\Tilde{p}_{\alpha} = \sum_{1\le i,j\le n} \Tilde{h}_{\alpha,i,j} \cdot \left((\sX\sY^\top)_{i,j} - \delta_{i,j}\right)$. Combining with \eqref{eq:p-alpha-ideal}, we finish the inductive step and conclude the proof.
\end{proof}

Next, we prove the existence of a special pseudoexpectation operator on $\bR[\sX,\sY]_{\le 2r}$. Our proof uses the following standard result about the feasibility of linear matrix inequalities.

\begin{lemma}[{\cite[Example 5.14]{vandenberghe2004convex}}] \label{lem:LMI-strong-alternative}
    Let $\FF_1, \dots, \FF_n$ and $\GG$ be real symmetric $k \times k$ matrices. Then, exactly one of the following holds:
    \begin{itemize}
        \item There exists $x_1, \dots, x_n \in \bR$ such that
        \begin{align*}
            \sum_{i=1}^n x_i \FF_i + \GG \succ 0.
        \end{align*}
        \item There exists a real symmetric matrix $\ZZ \succeq 0$ such that $\ZZ \ne \zzero$, $\tr(\FF_i \ZZ) = 0$ for all $1\le i \le n$, and $\tr(\GG \ZZ) \le 0$.
    \end{itemize}
\end{lemma}

\begin{proposition}\label{prop:pseudoexpectation}
    Let $r < n$ be positive integers. Let $\sX, \sY$ be $n \times r$ matrices of real indeterminates. There exists a pseudoexpectation operator $\Tilde{\bE}: \bR[\sX,\sY]_{\le 2r} \to \bR$ of degree $2r$, which satisfies:
    \begin{itemize}
        \item $\Tilde{\bE}$ is linear.
        \item $\Tilde{\bE}[1] = 1$.
        \item $\Tilde{\bE}[h\cdot ((\sX\sY^\top)_{i,j} - \delta_{i,j})] = 0$ for every $h \in \bR[\sX,\sY]$ of degree at most $2r-2$ and every $1\le i,j\le n$.
        \item $\Tilde{\bE}[p^2] \ge 0$ for every $p \in \bR[\sX,\sY]$ of degree at most $r$.
        \item Additionally, if $\Tilde{\bE}[p^2] = 0$ for some $p \in \bR[\sX,\sY]$ of degree at most $r$, then there exist polynomials $h_{i,j} \in \bR[\sX,\sY]$ for $1\le i,j\le n$ of degree at most $\deg(p) - 2$ such that
        \begin{align*}
            p = \sum_{1\le i,j\le n} h_{i,j} \cdot \left((\sX\sY^\top)_{i,j} - \delta_{i,j}\right).
        \end{align*}
    \end{itemize}
\end{proposition}

\begin{proof}
    For $0 \le d \le {2r}$, let
    \begin{align*}
        \calI_d \colonequals \left\{\sum_{1\le i,j\le n} h_{i,j} \cdot \left((\sX\sY^\top)_{i,j} - \delta_{i,j}\right): h_{i,j} \in \bR[\sX,\sY], \deg(h_{i,j}) \le d-2, \forall\, 1\le i,j\le n\right\}
    \end{align*}
    denote the space of polynomial combinations of $(\sX\sY^\top)_{i,j} - \delta_{i,j}$ of degree at most $d$. Note that $\calI_d$ is a linear subspace of $\bR[\sX,\sY]_{\le d}$. Moreover, since the constant polynomial $1$ does not vanish at $(\zzero, \zzero)$, $1 \not\in \calI_r$ by Proposition~\ref{prop:vanishing-structure}.
    
    Now we choose a basis of polynomials $\se_1, \dots, \se_K$ for $\calI_{r}$, and extend them to a basis of polynomials for $\bR[\sX,\sY]_{\le r}$ by choosing $\sv_1, \dots, \sv_N$ where $\sv_1 = 1$, so that $\se_1, \dots, \se_K, \sv_1, \dots, \sv_N$ form a basis of polynomials for $\bR[\sX,\sY]_{\le r}$. We denote $\sse = (\se_1, \dots, \se_K)$ and $\ssv = (\sv_1, \dots, \sv_N)$.

    We further define
    \begin{align*}
        \calL \colonequals \left\{L: \bR[\sX,\sY]_{\le 2r} \to \bR: L \text{ is linear}, \text{and }L(f) = 0, \forall\, f \in \calI_{2r}\right\}
    \end{align*}
    to be the collection of linear maps from the space of polynomials of degree at most $2r$ to the reals that vanish on the subspace $\calI_{2r}$. 
    
    For every linear map $L \in \calL$, we also associate to it a real symmetric matrix $\AA(L) \in \bR^{N \times N}$ whose entries are given by $\AA(L)_{i,j} = L(\sv_i\sv_j)$. Let $\calS$ denote the subspace of real symmetric $N \times N$ matrices spanned by $\AA(L)$ for $L \in \calL$. Since $\calS$ is finite dimensional, it is spanned by $\AA(L_1), \dots, \AA(L_D)$ for some $D \in \bN$ and $L_i \in \calL$ for $1\le i \le D$. By Lemma~\ref{lem:LMI-strong-alternative} applied to $\AA(L_1), \dots, \AA(L_D)$ with $\GG = \zzero$, either there exists an $N \times N$ positive definite matrix $\MM \in \calS$, or there exists an $N \times N$ real symmetric $\ZZ \succeq 0$ such that $\ZZ \ne \zzero$ and $\tr(\AA(L)\ZZ) = 0$ for every $L \in \calL$. We will show that the latter case is impossible through contradiction.

    Suppose for contradiction that there exists an $N \times N$ positive semidefinite matrix $\ZZ \ne \zzero$ such that $\tr(\AA(L)\ZZ) = 0$ for every $L \in \calL$. Write $\ZZ = \sum_{\alpha = 1}^m \cc_{\alpha} \cc_{\alpha}^\top$, where $m \ge 1$ and $\cc_{\alpha} \in \bR^N$ is nonzero for $1\le \alpha \le m$. We also define the polynomials $f_{\alpha} \colonequals \cc_{\alpha}^\top \ssv \in \bR[\sX,\sY]_{\le r}$. Then, for every $L \in \calL$,
    \begin{align*}
        0 &= \tr(\AA(L) \ZZ) = \tr\left(\sum_{\alpha = 1}^m \cc_{\alpha}^\top \AA(L) \cc_{\alpha}\right) =  L\left(\sum_{\alpha=1}^m\cc_{\alpha}^\top \ssv\ssv^\top \cc_{\alpha}\right) =  L\left(\sum_{\alpha=1}^m f_{\alpha}^2\right).
    \end{align*}
    Since $\calL$ is the subspace of linear maps that vanish on $\calI_{2r}$, it holds that $\sum_{\alpha=1}^m f_{\alpha}^2 \in \calI_{2r}$. By Proposition~\ref{prop:SoS-structure}, $f_{\alpha} \in \calI_{\deg(f_{\alpha})} \subseteq \calI_r$ for every $1\le \alpha \le m$. Recall that $\se_1, \dots, \se_K$ is a basis for $\calI_r$, and $\se_1, \dots, \se_K, \sv_1, \dots, \sv_N$ is a basis for $\bR[\sX,\sY]_{\le r}$. We see that $f_{\alpha} \in \calI_r$ contradicts that $\cc_{\alpha} \ne \zzero$, as a nonzero linear combination $f_{\alpha} = \cc_{\alpha}^\top \ssv$ of $\sv_1, \dots, \sv_N$ cannot be inside $\calI_r$.

    Hence, we have proved that there exists an $N \times N$ positive definite matrix $\MM \in \calS$. Note that $\MM = \AA(L_{\MM})$ for some $L_{\MM} \in \calL$. Now we define the pseudoexpectation operator $\Tilde{\bE}$ as follows. For every polynomial $f \in \bR[\sX,\sY]_{\le 2r}$, we set $\Tilde{\bE}[f] = \frac{L_{\MM}(f)}{L_{\MM}(1)}$, which is well-defined since $L_{\MM}(1) = \cc_1^\top \MM \cc_1 > 0$ where $\cc_1$ is the indicator vector of $\sv_1 = 1$. It is easy to see $\Tilde{\bE}$ is linear, evaluates to $1$ at $1$, and vanishes on $\calI_{2r}$. For a polynomial $p \in \bR[\sX,\sY]$ of degree at most $r$, if we write $p = \aa^\top \sse  +\cc^\top \ssv$, since $L_{\MM}$ vanishes on $\calI_{2r}$ , we have $L_{\MM}(p^2) =\cc^\top \MM \cc \ge 0$ and thus $\Tilde{\bE}[p^2] = \frac{L_{\MM}(p^2)}{L_{\MM}(1)} \ge 0$. If $\Tilde{\bE}[p^2] = 0$, since $\MM \succ 0$ is positive definite, it must hold that $\cc = 0$, i.e., $p = \aa^\top \sse \in \calI_{r}$, which implies $p \in \calI_{\deg(p)}$ by Proposition~\ref{prop:SoS-structure}.
\end{proof}

Given the special pseudoexpectation operator $\Tilde{\bE}$ supplied by Proposition~\ref{prop:pseudoexpectation}, now we are ready to prove that for large enough $B$, $\NormWRank_r^{n,B}(\II_n)$ does not have an SoS proof of degree at most $2r+1$.

\begin{proof}[Proof of Theorem~\ref{thm:norm-constraints-lower-bound}]
    Let $B > 0$ be a large enough constant to be set later.

    Suppose for contradiction that there exists an SoS proof of $\NormWRank_r^{n,B}(\II_n)$ of degree at most $2r+1$. Since $(\sX\sY^\top)_{i,j} - \delta_{i,j}$ and $\|\sX\|_F^2 + \|\sY\|_F^2 + \st^2 - B$ are unchanged when evaluated at $(-\sX,-\sY, -\st)$, taking the average of the polynomial identity of the SoS refutation at $(\sX,\sY, \st)$ and at $(-\sX,-\sY,-\st)$ cancels out the odd-degree components and gives an SoS proof of $\NormWRank_r^{n,B}(\II_n)$ of degree at most $2r$:
    \begin{align*}
        -1 &= \sum_{\alpha = 1}^m p_{\alpha}^2  + q \cdot (\|\sX\|_F^2 + \|\sY\|_F^2 + \st^2- B) + \sum_{1\le i,j \le n} h_{i,j} \cdot \left((\sX\sY^\top)_{i,j} - \delta_{i,j}\right), \numberthis \label{eq:SoS-proof-XYt}
    \end{align*}
    where $\deg(p_{\alpha}) \le r$, $\deg(q) \le 2r - 2$, and $\deg(h_{i,j}) \le 2r-2$ for every $1\le i,j\le n$.

    Next, we eliminate the indeterminate $\st$ as follows. For every $1\le \alpha \le m$, we write
    \begin{align*}
        p_{\alpha}(\sX,\sY,\st) \equalscolon \sum_{i=0}^{r} p_{\alpha,i}(\sX,\sY) \st^i,
    \end{align*}
    where $\deg(p_{\alpha,i}) \le r-i$. Averaging at $(\sX,\sY,\st)$ and at $(\sX,\sY,-\st)$ gives
    \begin{align*}
        \frac{p_{\alpha}(\sX,\sY,\st)^2 + p_{\alpha}(\sX,\sY,-\st)^2}{2} &= \frac{(p_{\alpha}(\sX,\sY,\st) + p_{\alpha}(\sX,\sY,-\st))^2 + (p_{\alpha}(\sX,\sY,\st) - p_{\alpha}(\sX,\sY,-\st))^2}{4}\\
        &= \left(\sum_{i=0}^{\lfloor r/2 \rfloor} p_{\alpha,2i}(\sX,\sY)\st^{2i}\right)^2 + \st^2\left(\sum_{i=0}^{\lfloor (r-1)/2 \rfloor} p_{\alpha,2i+1}(\sX,\sY)\st^{2i}\right)^2.
    \end{align*}
    Note that $\st$ only appears in the averaged polynomial of $p_{\alpha}$ as squares. Similarly, if we write $q \equalscolon \sum_{i=0}^{2r-2} q_{i}(\sX,\sY) \st^i$ and $h_{i,j} \equalscolon \sum_{k=0}^{2r-2} h_{(i,j),k}(\sX,\sY) \st^k$, averaging at $(\sX,\sY,\st)$ and at $(\sX,\sY,-\st)$ gives
    \begin{align*}
        \frac{q(\sX,\sY,\st) + q(\sX,\sY,-\st)}{2} &= \sum_{i=0}^{r-1} q_{2i}(\sX,\sY) \st^{2i}\\
        \frac{h_{i,j}(\sX,\sY,\st) + h_{i,j}(\sX,\sY,-\st)}{2} &= \sum_{k=0}^{r-1} h_{(i,j),2k}(\sX,\sY) \st^{2k},
    \end{align*}
    where again $\st$ only appears in the averaged polynomials as squares.

    Hence, if we average \eqref{eq:SoS-proof-XYt} at $(\sX,\sY,\st)$ and at $(\sX,\sY,-\st)$ and substitute $\st^2 = B - \|\sX\|_F^2 - \|\sY\|_F^2$, we get
    \begin{align*}
        -1 &= \sum_{\alpha = 1}^m \left(p_{\alpha, E}(\sX,\sY)^2 + (B - \|\sX\|_F^2 - \|\sY\|_F^2)p_{\alpha, O}(\sX,\sY)^2\right) + \sum_{1\le i,j\le n} \hat{h}_{i,j}(\sX,\sY) \cdot\left((\sX\sY^\top)_{i,j} - \delta_{i,j}\right), \numberthis \label{eq:SoS-proof-XY}
    \end{align*}
    where for $1\le \alpha \le m$,
    \begin{align*}
        p_{\alpha, E} &\colonequals \sum_{i=0}^{\lfloor r/2 \rfloor} p_{\alpha,2i}(\sX,\sY)(B - \|\sX\|_F^2 - \|\sY\|_F^2)^{i},\\
        p_{\alpha, O} &\colonequals \sum_{i=0}^{\lfloor (r-1)/2 \rfloor} p_{\alpha,2i+1}(\sX,\sY)(B - \|\sX\|_F^2 - \|\sY\|_F^2)^{i},
    \end{align*}
    which have degrees at most $r$ and $r-1$ respectively, and for $1\le i,j\le m$, 
    \begin{align*}
        \hat{h}_{i,j} \colonequals \sum_{k=0}^{r-1} h_{(i,j),2k}(\sX,\sY) (B - \|\sX\|_F^2 - \|\sY\|_F^2)^{k},
    \end{align*}
    which has degree at most $r-1$.

    By Proposition~\ref{prop:pseudoexpectation}, there exists a fixed pseudoexpectation operator $\Tilde{\bE}: \bR[\sX,\sY]_{2r} \to \bR$ independent of $B$, which is a linear map from the space of polynomials in $\sX$ and $\sY$ of degree at most $2r$ to the reals, such that 
    \begin{itemize}
        \item $\Tilde{\bE}[1] = 1$.
        \item $\Tilde{\bE}[h\cdot ((\sX\sY^\top)_{i,j} - \delta_{i,j})] = 0$ for every $h \in \bR[\sX,\sY]$ of degree at most $2r-2$ and every $1\le i,j\le n$.
        \item $\Tilde{\bE}[p^2] \ge 0$ for every $p \in \bR[\sX,\sY]$ of degree at most $r$.
    \end{itemize}
    Moreover, if $\Tilde{\bE}[p^2] = 0$ for some $p \in \bR[\sX,\sY]$ of degree at most $r$, then there exist polynomials $h_{i,j}^{(p)} \in \bR[\sX,\sY]$ for $1\le i,j\le n$ of degree at most $\deg(p) - 2$ such that
        \begin{align*}
            p = \sum_{1\le i,j\le n} h_{i,j}^{(p)} \cdot \left((\sX\sY^\top)_{i,j} - \delta_{i,j}\right).
        \end{align*}
    
    We use the standard fact that we may realize pseudoexpectation operators using positive semidefinite matrices as follows. Let $\ssv$ be a length-$N$ vector of all distinct monomials in $\sX,\sY$ of degree at most $r-1$. Then, for any polynomial $f \in \bR[\sX,\sY]$ of degree at most $r-1$, there exists a unique coefficient vector $\cc_f \in \bR^{N}$ such that $f(\sX,\sY) = \cc_f^\top \ssv$. Now consider two real symmetric matrices $\MM, \NN \in \bR^{N \times N}$ given by
    \begin{align*}
        \MM = \Tilde{\bE}[\ssv\ssv^\top] \qquad \text{ and } \qquad \NN = \Tilde{\bE}\left[(\|\sX\|_F^2 + \|\sY\|_F^2)\ssv\ssv^\top\right],
    \end{align*}
    which are well-defined as every entry of $\MM$ and $\NN$ is the evaluation of $\Tilde{\bE}$ on a polynomial of degree at most $2r$. Note that $\MM$ and $\NN$ are both positive semidefinite, since for any $\cc \in \bR^{N}$
    \begin{align*}
        \cc^\top \MM \cc &= \Tilde{\bE}[(\cc^\top \ssv)^2] \ge 0,\\
        \cc^\top \NN \cc &= \Tilde{\bE}[(\|\sX\|_F^2 + \|\sY\|_F^2)(\cc^\top \ssv)^2] \ge 0.
    \end{align*}
    
    Next, we show that whenever $\cc^\top \MM \cc = 0$, then $\cc^\top \NN \cc = 0$, i.e., $\ker(\MM) \subseteq \ker(\NN)$. Fix $\cc \in \bR^N$ such that $\cc^\top \MM \cc = 0$, and define $f(\sX,\sY) = \cc^\top \ssv$. Then, $\Tilde{\bE}[f^2] = \cc^\top \MM \cc = 0$. By the properties of the pseudoexpectation operator $\Tilde{\bE}$ given by Proposition~\ref{prop:pseudoexpectation}, there exist polynomials $h_{i,j}^{(f)} \in \bR[\sX,\sY]$ for $1\le i,j\le n$ of degree at most $\deg(f) - 2$ such that
        \begin{align*}
            f = \sum_{1\le i,j\le n} h_{i,j}^{(f)} \cdot \left((\sX\sY^\top)_{i,j} - \delta_{i,j}\right).
        \end{align*}
    Then, we have
    \begin{align*}
        \cc^\top \NN \cc &= \Tilde{\bE}\left[(\|\sX\|_F^2 + \|\sY\|_F^2) f^2\right]\\
        &= \Tilde{\bE}\left[(\|\sX\|_F^2 + \|\sY\|_F^2) f \sum_{1\le i,j\le n} h_{i,j}^{(f)} \cdot \left((\sX\sY^\top)_{i,j} - \delta_{i,j}\right)\right]\\
        &= 0,
    \end{align*}
    which vanishes due to the property of the pseudoexpectation operator $\Tilde{\bE}$, since $(\|\sX\|_F^2 + \|\sY\|_F^2) f h_{i,j}^{(f)}$ has degree at most $2 + \deg(f) + (\deg(f) - 2) = 2\deg(f) \le 2r-2$ for every $1\le i,j\le n$.

    Since $\MM$ and $\NN$ are fixed positive semidefinite matrices independent of $B$ and $\ker(\MM) \subseteq \ker(\NN)$, there exists a constant $C > 0$ independent of $B$ such that
    \begin{align*}
        \NN \preceq C \MM.
    \end{align*}
    In other words, for every polynomial $f \in \bR[\sX,\sY]$ of degree at most $r-1$, we have
    \begin{align*}
        \Tilde{\bE}\left[(\|\sX\|_F^2 + \|\sY\|_F^2)f^2\right] \le C\cdot \Tilde{\bE}[f^2]. \numberthis \label{ineq:C-pseudomoment-comparison}
    \end{align*}
    Fix such a constant $C > 0$. Applying the pseudoexpectation operator $\Tilde{\bE}$ to \eqref{eq:SoS-proof-XY}, we get
    \begin{align*}
        -1 &= \sum_{\alpha = 1}^m \left(\Tilde{\bE}[p_{\alpha, E}(\sX,\sY)^2] + B\cdot \Tilde{\bE}[p_{\alpha, O}(\sX,\sY)^2] -\Tilde{\bE}\left[(\|\sX\|_F^2 + \|\sY\|_F^2)p_{\alpha, O}(\sX,\sY)^2\right] \right)
    \intertext{By \eqref{ineq:C-pseudomoment-comparison}, we get}
        &\ge \sum_{\alpha = 1}^m \left(\Tilde{\bE}[p_{\alpha, E}(\sX,\sY)^2] + (B-C)\cdot \Tilde{\bE}[p_{\alpha, O}(\sX,\sY)^2] \right)
    \end{align*}
    which leads to a contradiction whenever $B \ge C$, since the $\Tilde{\bE}$ is nonnegative on squares. This finishes the proof that for a large enough $B = B(n,r) > 0$, $\NormWRank_r^{n,B}(\II_n)$ does not have an SoS proof of degree at most $2r+1$.
\end{proof}

\section{Proofs for Lower Bound Applications}

\subsection{Lower Bound for Sign Rank: Proof of Corollary~\ref{cor:sign-rank}}

We now derive Corollary~\ref{cor:sign-rank}, showing that refuting the system encoding that a sign matrix $\SS$ has sign rank at most $r$ needs an SoS degree at least $2r+2$.

\begin{proof}[Proof of Corollary~\ref{cor:sign-rank}]
    Assume we have a degree-$D$ SoS refutation of $\SRank_r^n(\SS)$:
    \begin{align*}
        -1 &= g + \sum_{i,j=1}^n h_{i,j} \cdot \left(\SS_{i,j} \sum_{k=1}^r \sX_{i,k}\sY_{j,k} - 1 - \sZ_{i,j}^2\right),
    \end{align*}
    where $g$ is a sum-of-squares polynomial in $\sX,\sY,\sZ$, and the displayed terms have degree at most $D$. Substituting $\sZ = \zzero$ in the identity above, we get
    \begin{align*}
        -1 = \hat{g} + \sum_{i,j=1}^n \SS_{i,j}\hat{h}_{i,j} \cdot \left( \sum_{k=1}^r \sX_{i,k}\sY_{j,k} - \SS_{i,j}\right),
    \end{align*}
    where $\hat{g}$ and $\hat{h}_{i,j}$ are obtained from $g$ and $h_{i,j}$ by substituting $\sZ=\zzero$. Note that this is a degree-$D$ SoS refutation of the system $\sX\sY^\top = \SS$. Since the real rank of $\SS$ is at least the sign rank of $\SS$, we conclude that $D \ge 2r+2$ by Theorem~\ref{thm:WRank-thm}.
\end{proof}

\subsection{Lower Bound for Matrix Rigidity: Proof of Corollary~\ref{cor:matrix-rigidity}}

We prove Corollary~\ref{cor:matrix-rigidity}, showing that certifying matrix non-rigidity inherits the SoS hardness of $\WRank$.

\begin{proof}[Proof of Corollary~\ref{cor:matrix-rigidity}]
    Assume we have a degree-$D$ SoS refutation of $\Rigid_r^s(\AA)$. If we substitute $\sZ = \sE = \zzero$ and $\st = \sqrt{s}$, then the original system becomes
    \begin{align*}
        \sX\sY^\top = \AA.
    \end{align*}
    The degree-$D$ SoS refutation of $\calR_{\AA}(r) \le s$ after this substitution becomes a degree-$D$ SoS refutation of $\sX\sY^\top = \AA$, and thus $\AA$ has rank at least $r+1$. By Theorem~\ref{thm:WRank-thm}, $D \ge 2r+2$.
\end{proof}

\subsection{Lower Bound for Algebraic Branching Programs: Proof of Corollary~\ref{cor:ABP}}

Now we use the SoS lower bound for $\WRank$ to deduce the SoS lower bound for refuting the existence of width-$r$ non-commutative ABP stated in Corollary~\ref{cor:ABP}.

\begin{proof}[Proof of Corollary~\ref{cor:ABP}]
    We use the reduction scheme from $\WRank$ to $\ncABP$ in \cite[Theorem~8.2]{garlik2026weak}. Fix a homogeneous non-commutative polynomial $f = \sum_{\substack{w: |w| = \ell}} c_w \cdot w \in \bR\langle z_1, \dots, z_n \rangle$ of degree $\ell$. Consider the following brute-force non-commutative ABP computing $f$: this ABP has layers $V_0, \dots, V_{\ell}$ where each internal layer $V_i$ for $1 \le i \le \ell - 1$ consists of $N \colonequals n^\ell$ vertices, and consists of $N$ paths from $s$ to $t$ that are vertex-disjoint at the internal layers. Each of the $N$ paths is identified with a word $w$ of length $\ell$, and computes the coefficient $c_w$ of the word $w$ in $f$. More precisely, for every $1\le i \le \ell-1$ and every word $w$ of length $\ell$, we create a vertex $(i, w) \in V_i$. The brute-force ABP computes $f$ as follows:
    \begin{itemize}
        \item Initialize $f_s = 1$.
        \item For every $|w| = \ell$, compute $f_{(1,w)} = f_s \cdot (c_w  w_1)$, where $w_1$ is the first letter of the word $w$.
        \item For every $2 \le i \le \ell-1$ and every $|w| = \ell$, compute $f_{(i,w)} = f_{(i-1,w)}w_i$, where $w_i$ is the $i$th letter of the word $w$.
        \item Compute $f_t = \sum_{\substack{w:|w| = \ell}} f_{(\ell-1, w)} w_{\ell}$, where $w_\ell$ is the last letter of the word $w$.
    \end{itemize}
    In matrix notation, it corresponds to the following computation of $f$ by the brute-force ABP:
    \begin{align*}
        f = M^{(1)}(\zz)M^{(2)}(\zz) \dots M^{(\ell)}(\zz),
    \end{align*}
    where $M^{(1)}(\zz)$ is a row vector of length $N$, $M^{(i)}(\zz)$ for $2 \le i \le \ell-1$ are $N \times N$ matrices, and $M^{(\ell)}(\zz)$ is a column vector of length $N$, whose entries are linear forms in $\zz$ given by
    \begin{align*}
        \hspace{0.6in}M^{(1)}_{s,w}(\zz) &= c_w w_1, && \forall\, |w| = \ell\\
        \hspace{0.6in}M^{(i)}_{w,w'}(\zz) &= \One\{w = w'\} w_{i}, && \forall\, |w| = |w'| = \ell\\
        \hspace{0.6in}M^{(\ell)}_{w,t}(\zz) &= w_{\ell}, && \forall\, |w| = \ell.
    \end{align*}

    If $N \le r$, padding this brute-force ABP with unused vertices gives a satisfying assignment to $\ncABP(f, r)$. Thus, we assume that $N > r$.

    For every $1\le i \le \ell$ and $1 \le j \le n$, we write
    \begin{align*}
        M^{(i)}(\zz) = \sum_{j=1}^n \MM^{(i,z_j)} z_j,
    \end{align*}
    where $\MM^{(i,z_j)}$ is the coefficient matrix of $z_j$ in $M^{(i)}(\zz)$ with real entries and the same dimension as $M^{(i)}(\zz)$. For every $1\le i \le \ell$ and every word $w$ of length $i$, we use $\pp^{(i,w)}$ to denote the coefficient row of $w$ in the prefix product $M^{(1)}(\zz)M^{(2)}(\zz) \dots M^{(i)}(\zz)$, i.e.,
    \begin{align*}
        \pp^{(i,w)} &\colonequals \MM^{(1,w_1)}\MM^{(2,w_2)} \dots \MM^{(i,w_i)},
    \end{align*}
    where $w_t$ is the $t$th letter of the word $w$. Equivalently, $\pp^{(i,w)}$ can be computed recursively as
    \begin{align*}
        \hspace{0.6in}\pp^{(1,z_j)} &= \MM^{(1,z_j)}, && \forall\, j \in [n]\\
        \hspace{0.6in}\pp^{(i,wz_j)} &= \pp^{(i-1,w)} \MM^{(i,z_j)}, && \forall\, 2 \le i \le \ell, j\in [n], |w| = i-1. 
    \end{align*}
    Since the brute-force ABP computes $f$, we have $\pp^{(\ell,wz_j)} = \pp^{(\ell-1,w)}\MM^{(\ell,z_j)}  = c_{wz_j}$ for every word $w$ of length $\ell-1$ and $j \in [n]$.

    Now assume we have a degree-$D$ SoS refutation of $\ncABP(f,r)$:
    \begin{align*}
        -1  &= g + \sum_{j=1}^n h^{(1,z_j)} \cdot (\sP^{(1,z_j)} - \sE^{(1,z_j)}) + \sum_{\substack{2 \le i \le \ell,\\ 1\le j \le n,\\ w: |w| = i-1} } h^{(i, wz_j)} \cdot (\sP^{(i,wz_j)} - \sP^{(i-1,w)}\sE^{(i,z_j)})\\
        &\hspace{0.4in}+ \sum_{w: |w| = \ell} q^{(w)} \cdot (\sP^{(\ell,w)} - c_w), \numberthis \label{eq:SoS-ABP}
    \end{align*}
    where $g$ is a sum-of-squares polynomial in $\sP^{(i,w)}$ and $\sE^{(i,z_j)}$, and all the displayed terms have degree at most $D$. Now let $\sX, \sY$ be two $N \times r$ matrices of real indeterminates, and consider the substitution 
    \begin{align*}
        \hspace{0.6in}\sE^{(1,z_j)} &= \MM^{(1,z_j)}\sX, &&\forall\, j\in [n]\\
        \hspace{0.6in}\sE^{(i,z_j)} &= \sY^\top \MM^{(i,z_j)} \sX, &&\forall\, 2\le i \le \ell-1, j\in [n]\\
        \hspace{0.6in}\sE^{(\ell,z_j)} &= \sY^\top \MM^{(\ell,z_j)}, &&\forall\, j\in [n]\\
        \hspace{0.6in}\sP^{(i,w)} &= \pp^{(i,w)}\sX, &&\forall\, 1\le i \le \ell-1, |w| = i\\
        \hspace{0.6in}\sP^{(\ell,w)} &= c_w, &&\forall\, |w| = \ell.
    \end{align*}
    Under this substitution, using the recursive formula for $\pp^{(i,w)}$, the SoS refutation \eqref{eq:SoS-ABP} becomes
    \begin{align*}
        -1  &= \hat{g} + \sum_{j=1}^n \hat{h}^{(1,z_j)} \cdot (\pp^{(1,z_j)} \sX - \MM^{(1,z_j)}\sX) + \sum_{\substack{2 \le i \le \ell-1,\\ 1\le j \le n,\\ w: |w| = i-1} } \hat{h}^{(i, wz_j)} \cdot (\pp^{(i,wz_j)}\sX - \pp^{(i-1,w)}\sX\sY^\top\MM^{(i,z_j)}\sX)\\
        &\hspace{0.4in}+ \sum_{\substack{1\le j \le n,\\ w: |w| = \ell-1} } \hat{h}^{(\ell, wz_j)} \cdot (c_{wz_j} - \pp^{(\ell-1,w)}\sX\sY^\top\MM^{(\ell,z_j)}) + \sum_{w: |w| = \ell} \hat{q}^{(w)} \cdot (c_w - c_w)\\
        &= \hat{g} + \sum_{\substack{2 \le i \le \ell-1,\\ 1\le j \le n,\\ w: |w| = i-1} } \hat{h}^{(i, wz_j)} \cdot (\pp^{(i-1,w)}(\II_N - \sX\sY^\top)\MM^{(i,z_j)}\sX)\\
        &\hspace{0.4in}+ \sum_{\substack{1\le j \le n,\\ w: |w| = \ell-1} } \hat{h}^{(\ell, wz_j)} \cdot (\pp^{(\ell-1,w)}(\II_N - \sX\sY^\top)\MM^{(\ell,z_j)}), \numberthis \label{eq:SoS-ABP-substitution}
    \end{align*}
    where $\hat{g}$ and $\hat{h}^{(i,w)}$ is obtained from $g$ and $h^{(i,w)}$ after the substitution of $\sP^{(i,w)}$ and $\sE^{(i,j)}$. Note that every substituted entry has degree at most $2$ in $\sX$ and $\sY$, and thus the polynomial $\hat{h}^{(i, wz_j)} \cdot (\pp^{(i-1,w)}(\II_N - \sX\sY^\top)\MM^{(i,z_j)}\sX)$ has degree at most $2(D-2) + 3 = 2D-1$ for every $2 \le i \le \ell=1$, $1\le j \le n$, and $|w| = i-1$, and similarly the polynomial $\hat{h}^{(\ell, wz_j)} \cdot (\pp^{(\ell-1,w)}(\II_N - \sX\sY^\top)\MM^{(\ell,z_j)})$ has degree at most $2D - 2$. Using the polynomial identity
    \begin{align*}
        \hat{g} &= -1 - \sum_{\substack{2 \le i \le \ell-1,\\ 1\le j \le n,\\ w: |w| = i-1} } \hat{h}^{(i, wz_j)} \cdot (\pp^{(i-1,w)}(\II_N - \sX\sY^\top)\MM^{(i,z_j)}\sX)\\
        &\hspace{0.4in} - \sum_{\substack{1\le j \le n,\\ w: |w| = \ell-1} } \hat{h}^{(\ell, wz_j)} \cdot (\pp^{(\ell-1,w)}(\II_N - \sX\sY^\top)\MM^{(\ell,z_j)}),
    \end{align*}
    we see that $\hat{g}$ also has degree at most $2D - 1$. Consequently, we conclude that \eqref{eq:SoS-ABP-substitution} is an SoS refutation of $\sX\sY^\top = \II_N$ of degree at most $2D-1$. By the SoS lower bound for $\WRank_r^N$ in Theorem~\ref{thm:WRank-thm}, we have $2D - 1 \ge 2r + 2$, which gives $D \ge r+2$.
\end{proof}

\subsection{Lower Bound for Tensor CP Rank: Proof of Corollary~\ref{cor:CPRank}}

We now derive an SoS lower bound for tensor CP rank using the previous SoS lower bound for non-commutative ABP.

\begin{proof}[Proof of Corollary~\ref{cor:CPRank}]
    Fix a $d$-tensor $\TT \in \bR^{n_1 \times \dots \times n_d}$. WLOG assume that $n_1 \ge n_2 \ge \dots \ge n_d$.

    Assume we have a degree-$D$ SoS refutation of $\CPRank_r^{(n_1, \dots, n_d)}(\TT)$: 
    \begin{align*}
        -1 &= g + \sum_{\substack{a_1, \dots, a_d:\\ a_i \in [n_i] \text{ for } 1\le i \le d} } h_{a_1, \dots, a_d} \cdot \left(\sum_{j=1}^r \prod_{i=1}^d \sV_{a_i,j}^{(i)} - \TT_{a_1, \dots, a_d}\right), \numberthis \label{eq:SoS-CP}
    \end{align*}
    where $g$ is a sum-of-squares polynomial in $\sV^{(i)}$ for $1\le i \le d$, and all the displayed term have degree at most $D$.
    
    Fix $2\le \ell \le d$. Let $N_{\ell} \colonequals \prod_{j=\ell+1}^{d} n_{j}$. Introduce formal variables $\zz^{(i)} = (z_1^{(i)}, \dots, z_{n_i}^{(i)})$ for $1 \le i \le \ell-1$ and $\yy = (y_{j,\bb})$ where $j \in [n_{\ell}]$ and $\bb = (b_1, \dots, b_{d-\ell})$ ranges over the $N_{\ell}$ possible grouped indices of the last $d - \ell$ tensor components. We associate to $\TT$ the following polynomial
    \begin{align*}
        f_{\TT}(\zz^{(1)}, \dots, \zz^{(\ell-1)}, \yy) = \sum_{\substack{a_1, \dots, a_{\ell}, \bb:\\ a_i \in [n_i] \text{ for } i \in [\ell],\\ \bb \in [n_{\ell+1}] \times \dots \times [n_d] }} \TT_{a_1, \dots, a_\ell, \bb} z_{a_1}^{(1)} \dots z_{a_{\ell-1}}^{(\ell-1)} y_{a_{\ell}, \bb}.
    \end{align*}

    The goal is to show that the SoS refutation of $\CPRank_r^{(n_1, \dots, n_d)}(\TT)$ above can be turned into an SoS refutation of $\ncABP(f,R)$ for some $R \ge 1$. Now we choose $R \ge 1$ such that $N_{\ell} R^{\ell-1} \le r$. Consider the subset $[N_{\ell} R^{\ell-1}] \subseteq [r]$ and identify this subset with the set of indices $$\{(\bb, \aalpha): \bb \in [n_{\ell+1}] \times \dots \times [n_d], \aalpha \in [R]^{\ell-1}\}.$$
    We denote $(R_0, R_1, \dots, R_{\ell-1}, R_{\ell}) = (1, R, \dots, R, 1)$. For $1 \le i \le \ell$ and a letter $z \in \bigcup_{i=1}^{\ell-1} \zz^{(i)} \cup \yy$, let $\sE^{(i,z)}$ be an $R_{i-1} \times R_i$ matrix of real indeterminates. Now consider the following substitution of the first $N_{\ell} R^{\ell-1}$ columns of $\sV^{(i)}$ for $1\le i \le d$, with the indices identified with $(\bb, \aalpha)$ as above. For a fixed $(\bb, \aalpha)$ where $\bb = (b_{\ell+1}, \dots, b_d) \in [n_{\ell+1}] \times \dots \times [n_d]$ and $\aalpha = (\alpha_1, \dots, \alpha_{\ell-1}) \in [R]^{\ell-1}$, substitute 
    \begin{align*}
        \hspace{0.4in}\sV^{(i)}_{j, (\bb, \aalpha)} &= \sE^{(i, z_j^{(i)})}_{\alpha_{i-1}, \alpha_i}, && \forall\, 1\le i \le \ell-1, 1 \le j \le n_i\\
        \hspace{0.4in}\sV^{(\ell)}_{j, (\bb, \aalpha)} &= \sE^{(\ell, y_{j,\bb})}_{\alpha_{\ell-1}, 1}, && \forall\, 1 \le j \le n_{\ell}\\
        \hspace{0.4in}\sV^{(i)}_{j,(\bb,\aalpha)} &= \delta_{j,b_i} && \forall\, \ell + 1 \le i \le d, 1 \le j \le n_i,
    \end{align*}
    where we denote $\alpha_0 \colonequals 1$, and we set the remaining $r - N_{\ell} R^{\ell-1}$ columns of $\sV^{(i)}$ to be zero. 

    Under this substitution, 
    \begin{align*}
        \sum_{j=1}^r \prod_{i=1}^d \sV_{a_i,j}^{(i)} &= \sum_{\substack{(\bb, \aalpha):\\ \bb \in [n_{\ell+1}] \times \dots \times [n_d],\\ \aalpha \in [R]^{\ell-1}} } \sE^{(1, z_{a_1}^{(1)})}_{1, \alpha_1}\sE^{(2, z_{a_2}^{(2)})}_{\alpha_{1}, \alpha_2} \dots \sE^{({\ell-1}, z_{a_{\ell-1}}^{(\ell-1)})}_{\alpha_{\ell-2}, \alpha_{\ell-1}}\sE^{(\ell, y_{a_{\ell}, \bb})}_{\alpha_{\ell-1}, 1} \prod_{i=\ell+1}^d\delta_{a_i, b_i}\\
        &= \sum_{\aalpha \in [R]^{\ell - 1}} \sE^{(1, z_{a_1}^{(1)})}_{1, \alpha_1}\sE^{(2, z_{a_2}^{(2)})}_{\alpha_{1}, \alpha_2} \dots \sE^{({\ell-1}, z_{a_{\ell-1}}^{(\ell-1)})}_{\alpha_{\ell-2}, \alpha_{\ell-1}}\sE^{(\ell, y_{a_{\ell}, (a_{\ell+1}, \dots, a_d)})}_{\alpha_{\ell-1}, 1}\\
        &= \sE^{(1, z_{a_1}^{(1)})}\sE^{(2, z_{a_2}^{(2)})} \dots \sE^{({\ell-1}, z_{a_{\ell-1}}^{(\ell-1)})}\sE^{(\ell, y_{a_{\ell}, (a_{\ell+1}, \dots, a_d)})},
    \end{align*}
    and the SoS refutation \eqref{eq:SoS-CP} becomes
    \begin{align*}
        -1 &= \hat{g} + \sum_{\substack{a_1, \dots, a_d:\\ a_i \in [n_i] \text{ for } 1\le i \le d} } \hat{h}_{a_1, \dots, a_d} \cdot \left(\sE^{(1, z_{a_1}^{(1)})} \dots \sE^{({\ell-1}, z_{a_{\ell-1}}^{(\ell-1)})}\sE^{(\ell, y_{a_{\ell}, (a_{\ell+1}, \dots, a_d)})} - \TT_{a_1, \dots, a_d}\right) \numberthis \label{eq:SoS-CP-ABP}
    \end{align*}
    where $\hat{g}$ and $\hat{h}_{a_1, \dots, a_d}$ are obtained from $g$ and $h_{a_1, \dots, a_d}$ after the substitution of the $\sV^{(i)}$ matrices. Since the substituted entries in the matrices have degree at most $1$ in $\sE^{(i,z)}$, all the polynomials
    $$\hat{h}_{a_1, \dots, a_d} \cdot \left(\sE^{(1, z_{a_1}^{(1)})} \dots \sE^{({\ell-1}, z_{a_{\ell-1}}^{(\ell-1)})}\sE^{(\ell, y_{a_{\ell}, (a_{\ell+1}, \dots, a_d)})} - \TT_{a_1, \dots, a_d}\right)$$
    have degree at most $D - (d - \ell)$. Consequently,
    $$\hat{g} = -1 - \sum_{\substack{a_1, \dots, a_d:\\ a_i \in [n_i] \text{ for } 1\le i \le d} } \hat{h}_{a_1, \dots, a_d} \cdot \left(\sE^{(1, z_{a_1}^{(1)})} \dots \sE^{({\ell-1}, z_{a_{\ell-1}}^{(\ell-1)})}\sE^{(\ell, y_{a_{\ell}, (a_{\ell+1}, \dots, a_d)})} - \TT_{a_1, \dots, a_d}\right)$$
    has degree at most $D - (d- \ell)$.

    Now for every $1 \le i \le \ell$ and every word $w$ of length $i$, we introduce a length-$R_i$ row vector $\sP^{(i,w)}$ of real indeterminates. Note that the following telescoping identity holds for any word $w$ of length $\ell$,
    \begin{align*}
        \sE^{(1, w_1)} \dots \sE^{(\ell, w_{\ell})} - c_w &= (\sP^{\ell,w} - c_w) - (\sP^{1,w_1} - \sE^{(1, w_1)})\prod_{t=2}^{\ell}\sE^{(t, w_t)} \\
        &\hspace{0.4in}- \sum_{i=2}^{\ell} (\sP^{i,w_1\dots w_i} - \sP^{i-1,w_1\dots w_{i-1}}\sE^{(i, w_i)})\prod_{t=i+1}^{\ell}\sE^{(t, w_t)}.
    \end{align*}
    Taking $w = z_{a_1}^{(1)}z_{a_2}^{(2)} \dots z_{a_{\ell-1}}^{(\ell-1)} y_{a_{\ell}, (a_{\ell+1}, \dots, a_d)}$ and $c_w = \TT_{a_1, \dots, a_d}$, the telescoping identity above turns \eqref{eq:SoS-CP-ABP} into a degree-$(D - (d-\ell))$ SoS refutation of $\ncABP(f_{\TT}, R)$. By Corollary~\ref{cor:ABP}, we have $D - (d - \ell) \ge R + 2$, and thus $D  \ge R + (d - \ell) + 2$.

    For fixed $2 \le \ell \le d$, the best $R$ under the constraint $N_{\ell} R^{\ell - 1} \le r$ is $R = \left\lfloor \left(\frac{r}{N_{\ell}}\right)^{\frac{1}{\ell-1}}\right\rfloor$. Taking the best bound among $2 \le \ell \le d$, we conclude that any SoS refutation of $\CPRank_r^{(n_1, \dots, n_d)}(\TT)$ needs degree at least
    \[ \max_{2\le \ell \le d} \left\lfloor \left(\frac{r}{N_{\ell}}\right)^{\frac{1}{\ell-1}}\right\rfloor + (d - \ell) + 2. \qedhere \]
\end{proof}

\appendix

\section{Proofs of Reductions}\label{app:proofs-of-reductions}

\subsection{Weak Rank, Tight Frame, and Pigeonhole Principles: Proof of Proposition~\ref{prop:proof-system-reduction}}\label{sec:proof-system-reduction}

\begin{proof}
    First, substitute $\sX\mapsto\AA\sX$ in a degree-$D$ refutation of $\sX\sY^\top=\AA$.
    Since
    \[
        (\AA\sX)\sY^\top-\AA=\AA(\sX\sY^\top-\II_n),
    \]
    each substituted constraint is a constant linear combination of the constraints for $\WRank_r^n(\II_n)$.
    The substitution preserves sums of squares and does not increase degree, so it gives a degree-$D$ refutation of $\sX\sY^\top=\II_n$.

    Next, setting $\sY=\sX$ in a refutation of $\sX\sY^\top=\II_n$ gives a refutation of $\sX\sX^\top=\II_n$.
    This substitution also preserves sums of squares and does not increase degree.

    Finally, each constraint for $\TFrame_r^n$ is a constant linear combination of the constraints for $\WPHP_r^n$:
    \begin{align*}
        (\sX\sX^\top)_{i,i}-1
        &=\sum_{j=1}^r(\sX_{i,j}^2-\sX_{i,j})
          +\left(\sum_{j=1}^r\sX_{i,j}-1\right),\\
        (\sX\sX^\top)_{i,k}
        &=\sum_{j=1}^r\sX_{i,j}\sX_{k,j} \text{ for } i\ne k.
    \end{align*}
    In a degree-$D$ refutation of $\TFrame_r^n$, each constraint has a multiplier of degree at most $D-2$.
    Replacing the constraints using these identities therefore gives a degree-$D$ refutation of $\WPHP_r^n$.
\end{proof}

\subsection{Inversion to Orthogonality: Proof of Proposition~\ref{prop:inversion-orthogonal-reduction}}\label{sec:inversion-orthogonal-reduction}

\begin{proof}
    Set $\sY=\sX$ in a degree-$D$ SoS proof of $\Inv_n$.
    The premise $\sX\sY^\top=\II_n$ becomes $\sX\sX^\top=\II_n$, and the conclusion $\sY^\top\sX=\II_n$ becomes $\sX^\top\sX=\II_n$.
    The substitution preserves sums of squares and does not increase degree, giving a degree-$D$ SoS proof of $\Orthog_n$.
\end{proof}

\subsection{Inversion to Weak Rank: Proof of Proposition~\ref{prop:inversion-rank-reduction}}\label{sec:inversion-rank-reduction}

\begin{proof}
    Let $\sX,\sY$ be $r \times r$ matrices of real indeterminates.
    Suppose there is a degree-$D$ SoS proof of $\Inv_{r}$:
    \begin{align}
        \sum_{i,j=1}^{r} (1 + k_{i,j}(\sX,\sY))\cdot\left((\sY^\top \sX)_{i,j} - \delta_{i,j}\right)^2 + g(\sX,\sY) = \sum_{i,j=1}^{r} h_{i,j}(\sX,\sY)\cdot\left((\sX\sY^\top)_{i,j}-\delta_{i,j}\right), \label{eq:inv-SoS}
    \end{align}
    where $g$ and each $k_{i,j}$ are sums of squares, and every displayed summand has degree at most $D$.

    Now consider two $(r+1) \times r$ matrices $\hat{\sX}, \hat{\sY}$ of real indeterminates. For $1 \le s \le r+1$, we let $\hat{\sX}^{(s)}$ and $\hat{\sY}^{(s)}$ denote the $r \times r$ submatrices obtained respectively from $\hat{\sX}$ and $\hat{\sY}$ by deleting the $s$th row.

    Substituting $\sX$ and $\sY$ in \eqref{eq:inv-SoS} with $\hat{\sX}^{(s)}$ and $\hat{\sY}^{(s)}$, we obtain for $1\le s \le r+1$,
    \begin{align*}
        &\sum_{i,j=1}^{r} (1 + k_{i,j}(\hat{\sX}^{(s)},\hat{\sY}^{(s)}))\cdot\left(((\hat{\sY}^{(s)})^\top \hat{\sX}^{(s)})_{i,j} - \delta_{i,j}\right)^2 + g(\hat{\sX}^{(s)},\hat{\sY}^{(s)}) \\
        &\hspace{0.4in}= \sum_{i,j=1}^{r} h_{i,j}(\hat{\sX}^{(s)},\hat{\sY}^{(s)})\cdot\left((\hat{\sX}^{(s)}(\hat{\sY}^{(s)})^\top)_{i,j}-\delta_{i,j}\right).
    \end{align*}
    Reexpressing the identity with $\hat{\sX}$ and $\hat{\sY}$, we get for $1 \le s\le r+1$,
    \begin{align*}
        &\sum_{i,j=1}^{r} (1 + k_{i,j}^{(s)}(\hat{\sX},\hat{\sY}))\cdot\left((\hat{\sY}^\top \hat{\sX})_{i,j} - \hat{\sY}_{s,i} \hat{\sX}_{s,j} - \delta_{i,j}\right)^2 + g^{(s)}(\hat{\sX},\hat{\sY}) \\
        &\hspace{0.4in}= \sum_{\substack{1\le i,j \le r+1:\\ i\ne s, j\ne s}} h_{i,j}^{(s)}(\hat{\sX},\hat{\sY})\cdot\left((\hat{\sX}\hat{\sY}^\top)_{i,j}-\delta_{i,j}\right),
    \end{align*}
    where we denote 
    \begin{align*}
        k_{i,j}^{(s)}(\hat{X},\hat{Y}) &\colonequals k_{i,j}(\hat{\sX}^{(s)},\hat{\sY}^{(s)}), \\ g^{(s)}(\hat{\sX},\hat{\sY}) &\colonequals g(\hat{\sX}^{(s)},\hat{\sY}^{(s)}), \\h_{i,j}^{(s)}(\hat{\sX}, \hat{\sY}) &\colonequals h_{i^{(s)},j^{(s)}}(\hat{\sX}^{(s)},\hat{\sY}^{(s)}),
    \end{align*} 
    with $i^{(s)} \colonequals i - \One\{i > s\}$. Summing over $1\le s \le r+1$, we get 
    \begin{align*}
        &\sum_{i,j=1}^{r+1} \left(\sum_{\substack{1 \le s \le r+1:\\ i\ne s,j\ne s}} h_{i,j}^{(s)}(\hat{\sX},\hat{\sY})\right)\left((\hat{\sX}\hat{\sY}^\top)_{i,j}-\delta_{i,j}\right)\\
        &= \sum_{s=1}^{r+1}\left(\sum_{i,j=1}^{r} (1 + k^{(s)}_{i,j}(\hat{\sX},\hat{\sY}))\cdot\left((\hat{\sY}^\top \hat{\sX})_{i,j} - \hat{\sY}_{s,i} \hat{\sX}_{s,j} - \delta_{i,j}\right)^2 + g^{(s)}(\hat{\sX},\hat{\sY})\right)\\
        &= \sum_{s=1}^{r+1}\left(g^{(s)}(\hat{\sX},\hat{\sY}) + \sum_{i,j=1}^r k^{(s)}_{i,j}(\hat{\sX},\hat{\sY}))\cdot\left((\hat{\sY}^\top \hat{\sX})_{i,j} - \hat{\sY}_{s,i} \hat{\sX}_{s,j} - \delta_{i,j}\right)^2 \right) \\
        &\hspace{0.4in} + \sum_{s=1}^{r+1}\sum_{i,j=1}^{r} \left((\hat{\sY}^\top \hat{\sX})_{i,j} - \hat{\sY}_{s,i} \hat{\sX}_{s,j} - \delta_{i,j}\right)^2
        \intertext{Note the first sum is a sum-of-squares polynomial of degree at most $D$. If we denote the first sum as $\hat{g}(\hat{\sX}, \hat{\sY})$ and expand the second sum, we get}
        &= \hat{g}(\hat{\sX}, \hat{\sY}) + (r+1)\sum_{i,j=1}^{r} \left((\hat{\sY}^\top \hat{\sX})_{i,j} - \delta_{i,j}\right)^2 + \sum_{i,j=1}^r \sum_{s=1}^{r+1} \hat{\sY}_{s,i}^2 \hat{\sX}_{s,j}^2 - 2\sum_{i,j=1}^r \sum_{s=1}^{r+1} \hat{Y}_{s,i} \hat{X}_{s,j}\left((\hat{\sY}^\top \hat{\sX})_{i,j} - \delta_{i,j}\right)
        \intertext{Using that that $\sum_{s=1}^{r+1} \hat{\sY}_{s,i} \hat{\sX}_{s,j} = (\hat{\sY}^\top \hat{\sX})_{i,j} = ((\hat{\sY}^\top \hat{\sX})_{i,j} - \delta_{i,j}) + \delta_{i,j}$, we get}
        &= \hat{g}(\hat{\sX}, \hat{\sY}) + (r-1)\sum_{i,j=1}^{r} \left((\hat{\sY}^\top \hat{\sX})_{i,j} - \delta_{i,j}\right)^2 + \sum_{s=1}^{r+1} \left(\sum_{i=1}^r\hat{\sY}_{s,i}^2\right) \left(\sum_{j=1}^r\hat{\sX}_{s,j}^2\right) - 2\sum_{i=1}^r \left((\hat{\sY}^\top \hat{\sX})_{i,i} - 1\right)
        \intertext{Note that $\sum_{i=1}^r (\hat{\sY}^\top \hat{\sX})_{i,i} = \tr(\hat{\sY}^\top \hat{\sX}) = \tr(\hat{\sX}\hat{\sY}^\top) = \sum_{s=1}^{r+1} (\hat{\sX}\hat{\sY}^\top)_{s,s}$, and by Lagrange's identity, $\left(\sum_{i=1}^r\hat{\sY}_{s,i}^2\right) \left(\sum_{j=1}^r\hat{\sX}_{s,j}^2\right) = \left(\sum_{i=1}^r \hat{\sX}_{s,i}\hat{\sY}_{s,i}\right)^2 + \sum_{1\le i < j \le r} \left(\sum_{i=1}^r \hat{\sX}_{s,i}\hat{\sY}_{s,j} - \hat{\sX}_{s,j}\hat{\sY}_{s,i}\right)^2$ Thus,}
        &= \hat{g}(\hat{\sX}, \hat{\sY}) + (r-1)\sum_{i,j=1}^{r} \left((\hat{\sY}^\top \hat{\sX})_{i,j} - \delta_{i,j}\right)^2 + \sum_{s=1}^{r+1} \left((\hat{\sX}\hat{\sY}^\top )_{s,s}^2 + \sum_{1\le i < j \le r} \left(\sum_{i=1}^r \hat{\sX}_{s,i}\hat{\sY}_{s,j} - \hat{\sX}_{s,j}\hat{\sY}_{s,i}\right)^2\right)\\
        &\hspace{0.4in}- 2\sum_{s=1}^{r+1} (\hat{\sX}\hat{\sY}^\top )_{s,s} + 2r
        \intertext{Completing the square, we get}
        &= \hat{g}(\hat{\sX}, \hat{\sY}) + (r-1)\sum_{i,j=1}^{r} \left((\hat{\sY}^\top \hat{\sX})_{i,j} - \delta_{i,j}\right)^2 + \sum_{s=1}^{r+1} \sum_{1\le i < j \le r} \left(\sum_{i=1}^r \hat{\sX}_{s,i}\hat{\sY}_{s,j} - \hat{\sX}_{s,j}\hat{\sY}_{s,i}\right)^2\\
        &\hspace{0.4in} +\sum_{s = 1}^{r+1} \left((\hat{\sX}\hat{\sY}^\top )_{s,s} - 1\right)^2 + (r-1).
    \end{align*}
    Rearranging the equation, we get
    \begin{align*}
        1 + G(\hat{\sX},\hat{\sY}) = \sum_{i,j=1}^{r+1}  H_{i,j}(\hat{\sX},\hat{\sY})\left((\hat{\sX}\hat{\sY}^\top)_{i,j}-\delta_{i,j}\right),
    \end{align*}
    where $G(\hat{\sX},\hat{\sY})$ is a sum-of-squares polynomial of degree at most $D$, and for $1\le i, j \le r+1$, \[H_{i,j}(\hat{\sX},\hat{\sY}) \colonequals \frac{1}{r-1}\sum_{\substack{1 \le s \le r+1:\\ i\ne s,j\ne s}} h_{i,j}^{(s)}(\hat{\sX},\hat{\sY})\]
    has degree at most $D-2$. This gives a degree-$D$ SoS proof of $\WRank_r^{r+1}(\II_{r+1})$.
\end{proof}

}

\addcontentsline{toc}{section}{References} 
\bibliographystyle{alpha}
\bibliography{main}

\end{document}